\documentclass[amsfonts,amssymb,amsfonts,amssymb,mathrsfs,graphicx,latexsym,url,amsthm,mathrsfs,amscd,verbatim,graphicx]{article} 
\usepackage{xcolor}
\usepackage{mathrsfs} 
\usepackage{graphicx,type1cm,eso-pic,color}
\usepackage{amsfonts, amssymb}

\newcommand{\qed}{\hfill $\square$}

\newcommand{\R}{{\mathbb R}}
\newcommand{\C}{{\mathbb C}}

\newcommand{\beq}{\begin{eqnarray}}
\newcommand{\eeq}{\end{eqnarray}}
\newcommand{\beqst}{\begin{eqnarray*}}
\newcommand{\eeqst}{\end{eqnarray*}}

\newcommand{\dsp}{\displaystyle}

\newtheorem{theorem}{Theorem}[section]
\newtheorem{lemma}[theorem]{Lemma}
\newtheorem{corollary}[theorem]{Corollary}
\newtheorem{proposition}{Proposition}[section]

\title{\bf 
Huygens' principle for the Dirac equation in  spacetime of non-constant curvature}

\author{{\bf Karen Yagdjian}}

\begin{document}

\date{}

\maketitle

\vspace*{-0.6cm}
\thispagestyle{empty}
\begin{center}
{\small  School of Mathematical and Statistical Sciences,}\\
{\small University of Texas RGV, 
1201 W.~University Drive,}  \\
{\small Edinburg, TX 78539,
USA} \\
{\small e-mail: karen.yagdjian@utrgv.edu} 
\end{center}
\medskip
\vspace*{-0.6cm}

\begin{abstract}
 In this article we give sufficient and necessary conditions for the validity of the  Huygens'  principle for   the  Dirac equation   in the non-constant curvature spacetime of the Friedmann-Lema\^itre-Robertson-Walker  models of cosmology. The   Huygens'  principle discussed for the   equation of a field with mass $m=0$ as well as a massive spin-$\frac{1}{2}$  field   undergoing a red shift  of its wavelength as the universe expands.

\medskip

\noindent
{\bf Keywords\,\, } Dirac equation $ \cdot $   Einstein~\&~de~Sitter model   $ \cdot $ FLRW   models  $ \cdot $   Huygens' principle $ \cdot $

\end{abstract}

\section{Introduction}

 In this article we give sufficient and necessary conditions for the validity of the  Huygens'  principle for   the  Dirac equation   in the non-constant curvature spacetime of the Friedmann-Lema\^itre-Robertson-Walker  models of cosmology.  
 We use the definition of the Huygens'  principle due to Hadamard \cite{Wunsch} as the absence of tails.  Thus,  
the field equations  
satisfy the Huygens'  principle if and only if the solution has no tail, that is, the solution depends
on the source distributions on the past null cone of the field only and not
on the sources inside the cone. 
\medskip

 The Dirac equation and its quantization in curved spacetime are of great
interest due to the role  of spin-$\frac{1}{2}$ particles in astrophysics and cosmology. Recent observational confirmation of the expansion of the  universe and  the quantum field theory  demand a detailed investigation of the solutions of the  Dirac equation in  curved spacetime (see, e.g., \cite{Birrell,Parker} and bibliography therein). 
The standard models of Cosmology provide such backgrounds,  which form a family of curved backgrounds of FLRW   models. 
For the  
de~Sitter spacetime in \cite{AP2020} a fundamental solution of the Dirac operator and an explicit formula for the solution of the Cauchy problem are  obtained. In \cite{JPHA2021} an examination  of these  explicit formulas gave an answer to such an interesting question in the physics of fundamental particles  as a validity of the Huygens' principle in the de~Sitter spacetime.
 It was proved in \cite{JMP2013} that the Klein-Gordon equation in the  de~Sitter universe obeys the Huygens' principle only for the particle with the mass   $\sqrt{2}$   (in the system of units with $c = \hbar = H=1$, where $c$ is the speed of light, $\hbar $
is the Planck's constant, and  $H$ is the Hubble constant), while  for the zero mass fields   the  incomplete Huygens' principle defined in  \cite{JMP2013} holds.  
In the case of a spatially curved spacetime  the last was revealed in \cite{Natario}.  
In fact, the fundamental solutions to the   operators 
can provide   information such as the Huygens' principle,  that is impossible to obtain by numerical solutions of differential equations. 
\medskip

For the  Dirac equation in a curved  four-dimensional spacetime,  
the Huygens'  principle is generally violated by its solutions, due to the  mass term in the  equation  and the curvature of spacetime \cite{Faraoni,Wunsch}.  The presence or absence of tails for   waves   has been established   for  some  spacetime metrics, including constant  curvature metrics \cite{Wunsch}. In fact, the study of  the Huygens'  principle has important applications to quantum field theory and cosmology,  especially in the inflationary theories of the early universe. 
The fact that  
the support of the commutator or the anticommutator-distribution, respectively,
lies on the null-cone if and only if the Huygens'  principle holds for the corresponding
 equation \cite{Lichnerowicz,Ooguri} shows significance of the Huygens'  principle    
for quantum field theory.   The Huygens' principle has been studied  in the
context of cosmology for classical fields \cite{Faraoni-Sonego} and for   gravitational waves in a  curved background (see, e.g.,   \cite{Kulczycki}).  
\medskip

For the Dirac operator in the cosmological context, the Huygens' principle  is discussed in 
\cite{McLenaghan2,Blasco2,Jonsson}.
The violations of
the  Huygens' principle \cite{Jonsson3}  may have  a consequence 
for relativistic quantum communication.
For the fields obeying the Huygens' principle communication through
massless fields is confined to the light cone. The violation of the Huygens' principle makes possible 
  a leakage of information towards the
inside of the light cone \cite{Jonsson3}. 
According to  \cite{Blasco2} the violation of the  Huygens' principle has unexpected
consequences 
in the propagation of information from the early
Universe to the current era 
\cite{Faraoni-Gunzig}. In \cite{Blasco2} are studied conformal and minimal couplings of a
test massless scalar field in a cosmological background.  The conclusion made in \cite{Blasco2} is that the 
signals received today 
  generically contains overlapped information about the
past from both timelike and light connected events. 
\medskip

In the theory of partial differential equations the   
Huygens'  principle has been used in the estimates of the Bahouri-G\'erard concentration 
compactness method and Strichartz estimates (see, e.g.,   \cite{Krieger,Schlag}) as well as in the study of global  solvability  of nonlinear hyperbolic equations (see, e.g.,  \cite{Alinhac}). 
\medskip

Our consideration is based on the  constructed in \cite{JDE_2021} the fundamental solution of the Dirac operator  and  the explicit formulas for the solution of the Cauchy problem  for the Dirac equation in the FLRW   spacetime with both accelerating and decelerating expansion or contraction. The spatially flat FLRW   models considered in \cite{JDE_2021} had 
the metric tensor that, in Cartesian coordinates, is written as follows 
\begin{equation}
\label{mtg}
(g_{\mu \nu })=    \left (
   \begin{array}{ccccc}
 1& 0& 0   & 0 \\
   0& -a^2(t) &  0 & 0 \\ 
 0 & 0 &  -a^2(t)   & 0 \\
 0& 0& 0   &  -a^2(t) \\
   \end{array} \right),\quad \mu ,\nu =0,1,2,3,  
\end{equation}
where the scale factor    $a(t)=a_0t^{\ell}  $, $\ell \in {\mathbb R}$,  $ t>0$,  and $x \in {\mathbb R}^3$, $x_0=t $. If $\ell<0$ the spacetime is contracting. In the case of $\ell>1$ the expansion is accelerating (with horizon), while for $0<\ell<1$ the expansion is decelerating. In the case of the Milne spacetime \cite{Sean Carroll,Gron-Hervik, Schrodinger}   $\ell=1 $. 
The FLRW  spacetimes with the scale factors   $a(t)=a_0t^{2/3}  $ and  $a(t)=a_0t^{1/2}  $   are modeling the matter dominated universe  and the radiation dominated universe, respectively (see, e.g., \cite{Ohanian-Ruffini}).  

It was proved by  W\"unsch~\cite{Wunsch} that if the massive ($m\not=0$) Dirac equation obeys the Huygens' principle, then  spacetime has a constant curvature.  Accordingly, it was admitted in \cite{JDE_2021} that the mass term of the field  can be changing in time and  vanishing at future infinity.
 The problem of the time variation of the  spin-$\frac{1}{2}$   massive particle in the early universe has been studied in physical literature 
(see \cite{Claudia} and the references therein).  In the present paper,   the decay assumption on the mass term has been made in the context of the expanding universe.  
 More exactly, the model is determined by the Dirac operator
\begin{equation}
\label{DO}
{\mathscr{D}}(t,\partial _t,\partial _x)
 := 
  \dsp 
 i {\gamma }^0    \partial_t   +i \frac{1}{a(t)}{\gamma }^1  \partial_{x_1}+i \frac{1}{a(t)}{\gamma }^2 \partial_{x_2}+i \frac{1}{a(t)}{\gamma }^ 3   \partial_{x_3} +i \frac{3\dot a(t)}{2a(t)}    {\gamma }^0     -m t^{-1}{\mathbb I}_4  \,,
\end{equation}
where $m \in {\mathbb C}$ and 
where 
the contravariant gamma matrices are  
\begin{eqnarray*}
&  &
 \gamma ^0= \left (
   \begin{array}{ccccc}
   {\mathbb I}_2& {\mathbb O}_2   \\
   {\mathbb O}_2& -{\mathbb I}_2   \\ 
   \end{array}
   \right),\quad 
\gamma ^k= \left (
   \begin{array}{ccccc}
  {\mathbb O}_2& \sigma ^k   \\
  -\sigma ^k &  {\mathbb O}_2  \\  
   \end{array}
   \right),\quad k=1,2,3\,.
 \end{eqnarray*}
Here $\sigma ^k $ are the Pauli matrices 
\begin{eqnarray*}
&  &
\sigma ^1= \left (
   \begin{array}{ccccc}
  0& 1   \\
  1& 0  \\  
   \end{array}
   \right), \quad
\sigma ^2= \left (
   \begin{array}{ccccc}
  0& -i   \\
  i& 0  \\  
   \end{array}
   \right),\quad
\sigma ^3= \left (
   \begin{array}{ccccc}
  1& 0   \\
  0&-1 \\  
   \end{array}
   \right)\,,
\end{eqnarray*}
and  ${\mathbb I}_n $, ${\mathbb O}_n $ denote the $n\times n$ identity and zero matrices, respectively. 

 This model includes the equation of a neutrino with $m=0$    as well as a massive spin-$\frac{1}{2}$ field  undergoing a red shifting of its wavelength as the universe expands. 
Thus, in the present paper we consider 
the Dirac equation in the   spacetime with the metric tensor (\ref{mtg}), that is,  
\[
{\mathscr{D}}(t,\partial _t,\partial _x)\Psi=F \,.
\]

 Recall  that   a retarded fundamental solution  for the Dirac operator (\ref{DO})  is a four-dimensional matrix  $ 
{\mathcal E}^{ret}={\mathcal E}^{ret} \left(x, t ; x_{0}, t_{0};m\right)
 $  with the 
operator-valued  entries      that solves the equation
\begin{equation}
\label{FSE} 
{\mathscr{D}}(t,\partial _t,\partial _x){\mathcal E}  \left(x, t ; x_{0}, t_{0};m\right) 
 = 
\delta ( x-x_0) \delta (t-t_0) {\mathbb I}_4, \qquad 
 (x,t), (x_0,t_0 ) \in {\mathbb R}^3\times {\mathbb R}_+,  
\end{equation}
and with the support in the {\it chronological future} (causal future) $D_+(x_0, t_0)$  of the point $(x_0,t_0)  \in {\mathbb R}^3\times {\mathbb R}_+$. The
advanced fundamental solution (propagator) $ 
{\mathcal E}^{adv}={\mathcal E}^{adv}  (x, t ; x_{0}, t_{0};$ $  m )
 $ solves the equation (\ref{FSE}) and has the  support in the {\it chronological past} (causal
past) $D_-(x_0, t_0)$.  The forward and backward light cones are defined as the boundaries of  
\[ 
D_{\pm}\left(x_{0}, t_{0}\right) :=\left\{(x, t) \in {\mathbb R}^3\times {\mathbb R}_+ \,;\,
\left|x-x_{0}\right| \leq \pm\left(\phi (t) -\phi (t_{0})  \right)\right\}\,,
\]
where $\phi (t):= \frac{1}{1-\ell}t^{1-\ell}$ if  $\ell \not=1$.   
In fact, any intersection of $D_-(x_0, t_0)$ with the hyperplane $0<t = const < t_0$ determines the
so-called {\it dependence domain} for the point $ (x_0, t_0)$, while the intersection of $D_+(x_0, t_0)$
with the hyperplane $t = const > t_0>0$ is the so-called {\it domain of influence} of the point
$ (x_0, t_0)$. 

\medskip

In \cite{JDE_2021} was defined   the right co-factor
\begin{eqnarray}
\label{Dcomp}
{\mathscr{D}}^{co} (t,\partial _t,\partial _x)
& := &
i t^{-\frac{\ell }{2}} \gamma ^0 (t^{ i m}  \gamma^U+t^{-i m}\gamma^L )\frac{\partial }{\partial t} 
+i t^{-\frac{3\ell }{2}}  \sum_{k=1}^3\gamma ^k (t^{ i m}  \gamma^U+t^{-i m}\gamma^L )\frac{\partial }{\partial x_k} 
\end{eqnarray}
 of the Dirac operator ${\mathscr{D}} (t,\partial _t,\partial _x) $ of (\ref{DO})  such that the composition   
${\mathscr{D}} (t,\partial _t,\partial _x){\mathscr{D}}^{co} (t,\partial _t,\partial _x) $ was a diagonal matrix of operators. Here, in order to distinguish upper and lower 2-spinors, 
 the upper-left corner and lower-right corner matrices,
\begin{eqnarray}
\label{gammaUL}
\gamma^U
& = &
\left(
\begin{array}{cc}
 {\mathbb I}_2 & {\mathbb O}_2    \\
 {\mathbb O}_2  & {\mathbb O}_2   \\ 
\end{array}
\right)=\frac{1}{2}({\mathbb I}_4+ \gamma^0 )
,\qquad 
 \gamma^L
  =  \left(
\begin{array}{cc}
 {\mathbb O}_2 & {\mathbb O}_2   \\
{\mathbb O}_2  & {\mathbb I}_2   \\ 
\end{array}
\right)=
\frac{1}{2}({\mathbb I}_4- \gamma^0 )\,,
\end{eqnarray}
respectively, will be used.

Let $\Delta  $ be the Laplace operator in ${\mathbb R}^3$. Denote  by ${\mathcal E}^w(x,t)$  the distribution that   
is the fundamental solution to the Cauchy problem for the  wave equation  in the Minkowski spacetime
\[
{\mathcal E}^w_{ tt} -   \Delta  {\mathcal E}^w  =  0 \,, \quad {\mathcal E}^w(x,0)=\delta (x)\,, \quad {\mathcal E}^w_{t}(x,0)= 0\,.
\]
According to Theorem~1.2~\cite{JDE_2021}, 
for every positive $ \varepsilon >0$ and $ t> \varepsilon $ the fundamental solution 
${\mathcal E}_{+}(x ,t;x_0 ;m;\varepsilon )$ to the Cauchy problem, that is, a distribution satisfying 
\begin{eqnarray*}
&  &
\cases{ {\mathscr{D}}(t,\partial _t,\partial _x)
{\mathcal E}_{+}(x ,t;x_0 ;m;\varepsilon )=   {\mathbb O}_4 ,  \cr
{\mathcal E}_{+}(x ,\varepsilon ;x_0 ;m;\varepsilon )=\delta ( x-x_0){\mathbb I}_4\,, }
\end{eqnarray*}
is given as follows
\begin{eqnarray*}
{\mathcal E}_{+}(x ,t;x_0 ;m;\varepsilon )
& = &
-i\varepsilon ^{1+\frac{\ell }{2} -i m} (1-\ell)^{-1}{\mathscr{D}}^{co} (x,t,\partial _t,\partial _x)\gamma^0\\
&  &
\times \int_0^{\phi (t)- \phi (\varepsilon )} \left (
   \begin{array}{ccccc}
K_1 \left(r,t; m ;\varepsilon \right)  {\mathbb I}_2& {\mathbb O}_2   \\
   {\mathbb O}_2& K_1 \left(r,t; -m ;\varepsilon \right) {\mathbb I}_2   \\ 
   \end{array}
   \right){\mathcal E}^w(x-x_0,r) \,dr     ,
\end{eqnarray*} 
where
\begin{eqnarray}
\label{K1def}
&  &
 K_1 \left(r,t; m ;\varepsilon \right)  \\
& := &
 2^{2 i  \frac{m}{1-\ell}} \phi (\varepsilon )^{2 i  \frac{m}{1-\ell}-1}   \left( \left(\phi (t)+ \phi (\varepsilon )  \right)^2-  r ^2\right)^{-  i \frac{m}{1-\ell}}   F \left(  i \frac{m}{1-\ell},  i \frac{m}{1-\ell};1;\frac{\left(\phi (t)- \phi (\varepsilon )\right)^2- r ^2}{\left(\phi (t)+ \phi (\varepsilon )   \right)^2-  r ^2 }\right)  \,. \nonumber 
\end{eqnarray} 
Henceforth, $F \left(  \alpha , \beta ;\gamma ;z\right) $ is the hypergeometric function (see, e.g., \cite{B-E}). In order to write  the solution to   the Cauchy problem  in an explicit form, 
  we use the integral operator 
\begin{eqnarray}
\label{K1oper}
&  &
{\cal K}_1(x,t,D_x;m;\varepsilon) [\varphi](x,t) \\
& := &
-i\varepsilon ^{1+\frac{\ell }{2} -i m}(1-\ell)^{-1}
\int_0^{ \phi (t)- \phi (\varepsilon ) }   K_1 \left(r,t; m ;\varepsilon \right)\int_{{\mathbb R}^3} {\mathcal E}^w(x-y,r)  \varphi (y )\,dy\,dr  
\,, \quad \varphi \in C_0^\infty({\mathbb R}^n). \nonumber
\end{eqnarray}
Theorem 1.3~\cite{JDE_2021} in the case of source free equation provides the representation of the solution to the Cauchy problem
\begin{eqnarray}
\label{EqAA}
\cases{{\mathscr{D}}(t,\partial _t,\partial _x)\Psi  (x,t)=0\,,\quad t>\varepsilon >0\,,\cr
\Psi (x,\varepsilon )=\Phi  (x)\,,}
\end{eqnarray}
with $m \in {\mathbb C}$,  as follows
\begin{equation}
\label{Sol}
\label{14}
\Psi  (x,t)
=
{\mathscr{D}}^{co} (t,\partial _t,\partial _x)\gamma^0\left (
   \begin{array}{ccccc}
  {\cal K}_1(x,t,D_x;m;\varepsilon ) {\mathbb I}_2& {\mathbb O}_2   \\
   {\mathbb O}_2&  {\cal K}_1(x,t,D_x;-m;\varepsilon ){\mathbb I}_2   \\ 
   \end{array}
   \right) [\Phi ]  (x,t)  ,\quad t> \varepsilon >0\,.
\end{equation} 

We say that the equation (\ref{EqAA}) obeys the     Huygens'  principle   if the solution $\Psi$  vanishes at all points which cannot be
reached from the support of initial data $\Phi  $ by a null geodesic.
The main result of this paper is the following theorem.
\begin{theorem}
\label{T0.3}
The solution of the Dirac equation  (\ref{EqAA}) with $\ell \in \R$, $\ell\not= 1$, and the mass $m \in {\mathbb C}$    obeys the Huygens'  principle if and only if the mass term takes the values $m=0, \pm i(\ell-1) $.  
\end{theorem}
According to Theorem~\ref{T0.3} for the spin-$\frac{1}{2}$ particle like neutrino,   the radiation and matter   dominated universes
are to some extent  opaque unless we accept that the mass of particle is either zero or   decaying in time.  Theorem~\ref{T0.3} guaranties that the Huygens' principle is fulfilled  and, consequently, the tail does not contribute to the dissipation of the the propagating field. In that sense for the     spin-$\frac{1}{2}$  massless or   decaying in time  imaginary  massive  fields with $m=\pm i(\ell-1) $, the   matter   dominated universe is transparent. 

 A discussion of an imaginary mass parameter in the Dirac equation from the physical point of view is given in \cite{JPHA2021}.

\smallskip
     
The rest of this paper is organized as follows.  In Section~\ref{S2}, we prove the sufficiency part of Theorem~\ref{T0.3} by  the representation formulas obtained in \cite{JDE_2021} for the solution of the   Dirac equation in the FLRW spacetime. In Section~\ref{S3} the  proof of the necessity 
part of Theorem~\ref{T0.3} is reduced to the verification of the  asymptotic (for large time) behavior of some integral. The asymptotic analysis is carried out  separately for the cases  $\ell >1$ and $\ell<1$ in Section~\ref{S4} and Section~\ref{S5}, respectively.

\section{Huygens' principle. Sufficient Conditions}
\label{S2}

The fundamental solution ${\mathcal E}^w$ to the Cauchy problem for the  wave equation  in the Minkowski spacetime can be written as ${\mathcal E}^w=\partial_t{\mathscr{V}}^w $, where
\[
{\mathscr{V}}^w_{ tt} -   \Delta  {\mathscr{V}}^w  =  0 \,, \quad {\mathscr{V}}^w(x,0)=0\,, \quad {\mathscr{V}}^w_{t}(x,0)= \delta (x)\,.
\]

In order to write the solution to   the Cauchy problem  for (\ref{EqAA}) 
  we use the operator  ${\cal K}_1(x,t,D_x;m;\varepsilon)$  defined  in (\ref{K1oper}).
According to Theorem 1.3~\cite{JDE_2021} in the case of source free equation, the solution to the Cauchy problem
(\ref{EqAA})
with $m \in {\mathbb C}$ is given by (\ref{14}). 
For $m=0$ we obtain
$
 K_1 \left(r,t; 0 ;\varepsilon \right) =
  \phi (\varepsilon )^{-1}$   
and, consequently,   
\begin{eqnarray*}
{\cal K}_1(x,t,D_x;0;\varepsilon) [\varphi](x,t) 
& := &
-i\varepsilon ^{\frac{\ell }{2} }(1-\ell)^{-1}
\int_0^{ \phi (t)- \phi (\varepsilon ) }    \int_{{\mathbb R}^3} {\mathcal E}^w(x-y,r)  \varphi (y )\,dy\,dr  
\,, \quad \varphi \in C_0^\infty({\mathbb R}^n).
\end{eqnarray*}
Hence the solution 
\[
{\cal K}_1(x,t,D_x;0;\varepsilon) [\varphi](x,t)=
-i\varepsilon ^{\frac{\ell }{2} }(1-\ell)^{-1}
  \int_{{\mathbb R}^3} {\mathscr{V}}^w(x-y,\phi (t)- \phi (\varepsilon ))  \varphi (y )\,dy 
\,, \quad \varphi \in C_0^\infty({\mathbb R}^n),
\]
obeys the Huygens' principle.
\medskip

For the case of  the values of  mass $m=\pm  i(1-\ell)$ we have 
\begin{eqnarray*}
 K_1 \left(r,t; i(1-\ell) ;\varepsilon \right) 
& = &
r^2\left( \frac{1}{2} \ell^3  \varepsilon ^{3 \ell-3}-\frac{3}{2} \ell^2  \varepsilon ^{3 \ell-3}-\frac{1}{2}  \varepsilon ^{3 \ell-3}+\frac{3}{2} \ell  \varepsilon ^{3 \ell-3} \right)\\
&  &
+ \left( \frac{1}{2} \varepsilon ^{3\ell-3} t^{2-2\ell}-\frac{1}{2} \ell \varepsilon ^{3\ell-3} t^{2-2\ell}-\frac{1}{2} \ell \varepsilon ^{\ell-1}+\frac{\varepsilon ^{\ell-1}}{2} \right),\\
K_1 \left(r,t; -i(1-\ell) ;\varepsilon \right)  
& = &
-(\ell-1) t^{\ell-1}\,.
\end{eqnarray*} 
In the next theorem we consider the family of   more general   kernels $K_1 \left(r,t; \pm m ;\varepsilon \right) $, which leads to the Huygens' principle.  
\begin{theorem}
\label{T2.1}
If the kernels $ K_1 \left(r,t; \pm m ;\varepsilon \right) $ can be represented in the form
\begin{eqnarray*}
&  &
 K_1 \left(r,t; \pm m ;\varepsilon \right) =r^2 a_\pm \left( \pm m ;\varepsilon \right)+b_\pm\left(t; \pm m ;\varepsilon \right)\,,
\end{eqnarray*}
then the Dirac equation (\ref{EqAA}) obeys the Huygens' principle.   
\end{theorem}
\medskip

\noindent
{\bf Proof.} 
First, we consider the case of plus. According to (\ref{K1oper})  we have
\begin{eqnarray}
\label{11}
&  &
{\cal K}_1(x,t,D_x;m;\varepsilon) [\varphi](x,t) \\
& = &
-i\varepsilon ^{1+\frac{\ell }{2} -i m}(1-\ell)^{-1}a_+ \left(   m ;\varepsilon \right)
\int_0^{ \phi (t)- \phi (\varepsilon ) }    r^2\int_{{\mathbb R}^3} {\mathcal E}^w(x-y,r)  \varphi (y )\,dy\,dr \nonumber \\
&   &
-i\varepsilon ^{1+\frac{\ell }{2} -i m}(1-\ell)^{-1}b_+\left(t;   m ;\varepsilon \right)
\int_0^{ \phi (t)- \phi (\varepsilon ) }    \int_{{\mathbb R}^3} {\mathcal E}^w(x-y,r)  \varphi (y )\,dy\,dr 
\,, \quad \varphi \in C_0^\infty({\mathbb R}^n). \nonumber 
\end{eqnarray}
For the last term of the previous relation we have
\[ 
\int_0^{ \phi (t)- \phi (\varepsilon ) }  \int_{{\mathbb R}^3} {\mathcal E}^w(x-y,r)  \varphi (y )\,dy\,dr
=
  \int_{{\mathbb R}^3} {\mathscr{V}}^w(x-y,\phi (t)- \phi (\varepsilon ))  \varphi (y )\,dy 
\,, \quad \varphi \in C_0^\infty({\mathbb R}^n).
\]
This term obeys the Huygens' principle since ${\mathscr{V}}^w $ does it. Next  we consider the first term of that relation (\ref{11}):
\begin{eqnarray*}
&  &
\int_0^{ \phi (t)- \phi (\varepsilon ) }  r^2\int_{{\mathbb R}^3} {\mathcal E}^w(x-y,r)  \varphi (y )\,dy\,dr\\
& = &
\left( r^2\int_{{\mathbb R}^3} {\mathscr{V}}^w(x-y,r)  \varphi (y )\,dy\right)_{r=\phi (t)- \phi (\varepsilon )}
-  2\int_0^{ \phi (t)- \phi (\varepsilon ) }r  \left( \int_{{\mathbb R}^3} {\mathscr{V}}^w(x-y,r)  \varphi (y )\,{d}y\right)\,{d}r\,.
\end{eqnarray*} 
In view of (\ref{14}) and the structure of ${\mathscr{D}}^{co} (t,\partial _t,\partial _x) $, the following lemma completes the proof of Theorem~\ref{T2.1}.
\begin{lemma}
The function 
\begin{eqnarray*}
\frac{\partial}{\partial x_j}\int_0^{ \phi (t)- \phi (\varepsilon ) }   r   \left( \int_{{\mathbb R}^3} {\mathscr{V}}^w(x-y,r)  \varphi (y )\,dy\right)\,{d} r 
\,, \quad \varphi \in C_0^\infty({\mathbb R}^3),\quad j=1,2,3,
\end{eqnarray*}
satisfies the Huygens' principle. 
\end{lemma}
\medskip

\noindent
{\bf Proof.} Indeed, we apply the Kirchhoff's formula and consider, for instance, the case of $j=3$. Then up to an unimportant factor, the possible tail is:
\begin{eqnarray*}
&  &
\frac{\partial}{\partial x_j}\int_0^{ \phi (t)- \phi (\varepsilon ) }   r   \left( \int_{{\mathbb R}^3} {\mathscr{V}}^w(x-y,r)  \varphi (y )\,dy\right)\,{d} r \\
& = &
\int\!\int\!\int_{|y| \leqslant \phi (t)- \phi (\varepsilon )} \frac{\partial}{\partial y_{3}} \varphi (x+y) {d} y_{1} {~d} y_{2} {~d} y_{3} \\
& = & 
\int_{y_{1}^{2}+y_{2}^{2} \leqslant (\phi (t)- \phi (\varepsilon ))^{2}}\left\{\varphi \left(x_{1}+y_{1}, x_{2}+y_{2}, x_{3}+\sqrt{(\phi (t)- \phi (\varepsilon ))^{2}-y_{1}^{2}-y_{1}^{2}}\right)\right.\\
&  &
\left.-\varphi \left(x_{1}+y_{1}, x_{2}+y_{2}, x_{3}-\sqrt{(\phi (t)- \phi (\varepsilon ))^{2}-y_{1}^{2}-y_{1}^{2}}\right)\right\} {d} y_{1} {~d} y_{2}\,.
\end{eqnarray*}
For every $t>\varepsilon $ the points
\[
\left(x_{1}+y_{1}, x_{2}+y_{2}, x_{3} \pm \sqrt{(\phi (t)- \phi (\varepsilon ))^{2}-y_{1}^{2}-y_{1}^{2}}\right)
\in \R^3, \quad \mbox{\rm where}\quad y_{1}^{2}+y_{2}^{2} \leqslant (\phi (t)- \phi (\varepsilon ))^{2}\,,
\]
belong to the sphere of the radius $(\phi (t)- \phi (\varepsilon ))$    in $\R^3$, that is, the domain of integration does not intersect
the interior of the domain of dependence. Thus, the tail is empty and the theorem is proved.
\qed

\section{Huygens' principle. The necessary conditions}
\label{S3}

The proof of the necessity part will be carried out in two steps. The first step is the choice of the special initial spinor that is a radial spinor with   a support in small neighborhood of the origin.  The second step is in the establishing asymptotic behavior of the solution for large time at the spatial origin. If the value of the solution  at the spatial origin differs from zero for the large time, then the Huygens' principle is violated.

According to Theorem~1.3~\cite{JDE_2021}, 
the solution to the Cauchy problem (\ref{EqAA}) 
with $m \in {\mathbb C}$, is given by (\ref{Sol}),  
where  the right co-factor $ {\mathscr{D}}^{co} (t,\partial _t,\partial _x)$ is given by (\ref{Dcomp}), while $\gamma^U $ and $\gamma^L $ are defined in (\ref{gammaUL}). 
The proof of the necessity part of  Theorem~\ref{T0.3} is based on the large time asymptotics  of the tail of solution.  The initial data $ \Phi (x) = ( \Phi _0(x ) , \Phi _1(x ), \Phi _2(x ), \Phi _3(x ))^T$ will be chosen radial
having small support. Consider the solution  of the Cauchy problem with the radial  function $\Phi  (x )=\Phi   (r ) $,
supp$\,\Phi   \subset \left\{x \in {\mathbb R}^{n} ;|x| \leq \min\{ 1/3,  {\widetilde\varepsilon} /|\ell -1| \}  \right\}$, ${\widetilde\varepsilon}  \in (0,1)$:
\begin{eqnarray*}
\Psi  (x,t)
& = &
 {\mathscr{D}}^{co} (t,\partial _t,\partial _x)\gamma^0\left (
   \begin{array}{cccc}
 {\cal K}_1(x,t,D_x;m;\varepsilon )[\Phi _0(x ) ]  \\
{\cal K}_1(x,t,D_x;m;\varepsilon )[ \Phi _1(x )] \\
 {\cal K}_1(x,t,D_x;-m;\varepsilon )[\Phi _2(x )]  \\
 {\cal K}_1(x,t,D_x;-m;\varepsilon )[\Phi _3(x ) ]\\
   \end{array}
   \right )  (x,t)  ,\quad t> \varepsilon >0\,.
\end{eqnarray*} 
If we choose the initial data 
\[
\Phi (x) = ( \Phi _0(x ) ,0,0,0)^T\,,
\] 
 then the solution $\Psi  (x,t) = ( \Psi_0  (x,t) ,\Psi_1 (x,t) ,\Psi_2 (x,t) ,\Psi_3  (x,t))^T $  is given by
 \begin{eqnarray*}
\Psi  (x,t)
& = &
\left( i t^{-\frac{\ell }{2}} \gamma ^0 (t^{ i m}  \gamma^U+t^{-i m}\gamma^L )\frac{\partial }{\partial t} 
+i t^{-\frac{3\ell }{2}}  \sum_{k=1}^3\gamma ^k (t^{ i m}  \gamma^U+t^{-i m}\gamma^L )\frac{\partial }{\partial x_k}\right)\\
&  &
\times \gamma^0\left (
   \begin{array}{cccc}
 {\cal K}_1(x,t,D_x;m;\varepsilon )[\Phi _0(x ) ]  \\
0 \\
0  \\
0\\
   \end{array}
   \right )  (x,t)  ,\quad t> \varepsilon >0\,.
\end{eqnarray*} 
The first component of $\Psi  (x,t) $ is 
\begin{eqnarray*}
&  &
\Psi_0  (x,t)\\
& = &
i t^{-\frac{\ell }{2}+ i m} \frac{\partial }{\partial t} 
\left (
 {\cal K}_1(x,t,D_x;m;\varepsilon )[\Phi _0(x ) ] 
   \right )  (x,t)\\
& = &
 t^{-\frac{\ell }{2}+ i m}\varepsilon ^{1+\frac{\ell }{2} -i m}(1-\ell)^{-1} \frac{\partial }{\partial t} 
\left (
\int_0^{ \phi (t) - \phi (\varepsilon )}   K_1 \left(r,t; m ;\varepsilon \right)\int_{{\mathbb R}^3} {\mathcal E}^w(x-y,r)  \Phi _0 (y )\,dy\,dr 
   \right )  ,\quad t> \varepsilon >0\,.
\end{eqnarray*}
If we choose the initial data
\[
\Phi (x) = ( 0 ,0,\Phi _2(x ),0)^T\,,
\]
 then the solution $\Psi  (x,t) = ( \Psi_0   (x,t),\Psi_1  (x,t),\Psi_2  (x,t),\Psi_3  (x,t))^T $  is given by
 \begin{eqnarray*}
\Psi  (x,t)
& = &
\left( i t^{-\frac{\ell }{2}} \gamma ^0 (t^{ i m}  \gamma^U+t^{-i m}\gamma^L )\frac{\partial }{\partial t} 
+i t^{-\frac{3\ell }{2}}  \sum_{k=1}^3\gamma ^k (t^{ i m}  \gamma^U+t^{-i m}\gamma^L )\frac{\partial }{\partial x_k}\right)\\
&  &
\times \gamma^0\left (
   \begin{array}{cccc}
0   \\
0 \\
{\cal K}_1(x,t,D_x;m;\varepsilon )[\Phi _2(x ) ]  \\
0\\
   \end{array}
   \right )  (x,t)  ,\quad t> \varepsilon >0\,.
\end{eqnarray*} 
The third component of $ \Psi  (x,t)$ is 
\begin{eqnarray*}
&  &
\Psi_2  (x,t)\\
& = &
- i t^{-\frac{\ell }{2}-i m}  \frac{\partial }{\partial t} 
{\cal K}_1(x,t,D_x;-m;\varepsilon )[\Phi _2(x ) ](x,t)\\
& = &
-  \varepsilon ^{1+\frac{\ell }{2} +i m}(1-\ell)^{-1}t^{-\frac{\ell }{2}-i m}  \frac{\partial }{\partial t} 
\Bigg( 
\int_0^{ \phi (t)- \phi (\varepsilon ) }   K_1 \left(r,t; -m ;\varepsilon \right)\int_{{\mathbb R}^3} {\mathcal E}^w(x-y,r) \Phi _2(y )\,dy\,dr \Bigg)    ,\,\, t> \varepsilon >0\,.
\end{eqnarray*} 

Denote either $\varphi (y ):= \Phi _0(y ) $ or $\varphi (y ):= \Phi _2(y ) $. We need to find a large time  asymptotics only for 
the integral
\begin{eqnarray*}
&  & 
\int_0^{ \phi (t)- \phi (\varepsilon ) } \left( \frac{\partial }{\partial t}   K_1 \left(r,t; \pm m ;\varepsilon \right)
\right)\int_{{\mathbb R}^3} {\mathcal E}^w(x-y,r)  \varphi (y )\,dy\,dr \,,
\end{eqnarray*}
since the term obtained by differentiation of the upper limit, 
\begin{eqnarray*}
&  & 
t^{-\ell}  K_1 \left(\phi (t)- \phi (\varepsilon ),t; \pm m ;\varepsilon \right)\int_{{\mathbb R}^3} {\mathcal E}^w(x-y,\phi (t)- \phi (\varepsilon ))  \varphi (y )\,dy \,,
\end{eqnarray*}
obeys the Huygens' principle. Hence, we consider
\begin{eqnarray*}
&  & 
\int_0^{ \phi (t)- \phi (\varepsilon ) } \left( \frac{\partial }{\partial t}   K_1 \left(r,t; \pm m ;\varepsilon \right)
\right) \partial_r\int_{{\mathbb R}^3} {\mathscr{V}}^w(x-y,r)  \varphi (y )\,dy\,dr\,.
\end{eqnarray*}
Denote
\[
\widetilde \Psi (x,t)
  :=   
\int_0^{ \phi (t)- \phi (\varepsilon ) } \left( \frac{\partial }{\partial t}   K_1 \left(r,t; \pm m ;\varepsilon \right)
\right) \partial_r{\mathscr{V}}_\varphi  (x,r)\,dr\,,
\]
where the notation
\[
{\mathscr{V}}_\varphi  (x,r):=\int_{{\mathbb R}^3} {\mathscr{V}}^w(x-y,r)  \varphi (y )\,dy\,
\]
has been used. Then
\begin{eqnarray*}
\widetilde \Psi (x,t)
& = & 
  \left( \frac{\partial }{\partial t}   K_1 \left( \phi (t)- \phi (\varepsilon ),t; \pm m ;\varepsilon \right)
\right)  {\mathscr{V}}_\varphi  (x, \phi (t)- \phi (\varepsilon ))\,dr\\
&  &
-\int_0^{ \phi (t)- \phi (\varepsilon ) } \left( \frac{\partial }{\partial r} \frac{\partial }{\partial t}   K_1 \left(r,t; \pm m ;\varepsilon \right)
\right) {\mathscr{V}}_\varphi  (x,r)\,dr\,.
\end{eqnarray*}
In particular,  by the Kirchhoff's formula we have ${\mathscr{V}}_\varphi   (0, r) = r\varphi  (r)$ and
\begin{eqnarray*}
{\mathscr{V}}_\varphi (0, \phi (t)- \phi (\varepsilon ))=  (\phi (t)- \phi (\varepsilon )) \varphi  _0 (\phi (t)- \phi (\varepsilon )) =0 
\end{eqnarray*} 
for sufficiently large $t$, that is, if $\phi (t)- \phi (\varepsilon ) >{\widetilde\varepsilon} $. 
Consequently, for large $t$ we have
\begin{eqnarray}
\widetilde \Psi (0,t)
& = & 
-\int_0^{ \phi (t)- \phi (\varepsilon ) } \left( \frac{\partial }{\partial r} \frac{\partial }{\partial t}   K_1 \left(r,t; \pm m ;\varepsilon \right)
\right) r\varphi  (r)\,dr \nonumber \\
\label{16}
& = & 
\int_0^{ \widetilde\varepsilon  } \left(\frac{\partial }{\partial t}   K_1 \left(r,t; \pm m ;\varepsilon \right)
\right)  \frac{\partial }{\partial r} (r\varphi  (r))\,dr\,.
\end{eqnarray}

The outline of the remaining part of the proof is as follows. In the next sections we study the asymptotcs of the function $\frac{\partial }{\partial t}   K_1 \left(r,t; \pm m ;\varepsilon \right) $ as $t \to \infty $. The principal term of this  asymptotics for the cases of $m\not= 0,\pm i(\ell-1)$ is the function of  $r$  that allows us to find the initial function $\varphi  (r) $ such that the scalar product (\ref{16}) of $ \frac{\partial }{\partial r} (r\varphi  (r))$ with the principal term differs from zero. Thus, $\widetilde \Psi (0,t) \not=0$ for sufficiently large time, and  the Huygens' principle is violated for these values of $m$. 

It is  easily  seen that it  suffices to study the asymptotic behavior  of the function 
\[
\frac{\partial}{\partial \tau } \left(\left((\tau +1)^2-A^2\right)^{-M} F\left(M,M;1;\frac{(\tau -1)^2-A^2}{(\tau +1)^2-A^2}\right)\right)
\]
for $M \in \C$, $ M\not=0,\pm 1$, as $\tau \to 0$ if $\ell>1 $ or as $\tau \to \infty$ if $\ell<1 $. For $M=0,\pm 1$ the Huygens' principle holds. Here $\tau =t^{1-\ell}$ while $A^2:= (\ell -1)^2 r^2$ is sufficiently small, say $A^2 \leq 1/9$. It is important that 
\[
0< \frac{(\tau -1)^2-A^2}{(\tau +1)^2-A^2}<1\quad \mbox{\rm when}\quad \tau \in (0,1/2)\cup (2,\infty)\quad \mbox{\rm and }\quad A^2 \in[0, 1/9],
\]
as well as  
\[
\lim_{ \tau \to 0}  \frac{(\tau -1)^2-A^2}{(\tau +1)^2-A^2}=\lim_{ \tau \to \infty}  \frac{(\tau -1)^2-A^2}{(\tau +1)^2-A^2} = 1\quad \mbox{\rm uniformly on}\quad   A^2 \in [0,1/9]\,.
\]
Moreover, we will use the finite sum 
\begin{eqnarray*}
&  &
\sum_{k=N_1}^{N_2}\tau ^{a_ k}\left( h_k(A,M)+{\widetilde h}_k(A,M) (\ln (\tau ))^{b_k} \right)
\end{eqnarray*} 
of the terms of the  asymptotic   (as $t \to \infty$) series 
\begin{eqnarray*}
&  &
\sum_{k=-\infty}^{\infty}\tau ^{ a_k}\left( h_k(A,M)+{\widetilde h}_k(A,M) (\ln (\tau ))^{b_k} \right)\,,
\end{eqnarray*} 
where $ a_k \in \C$, $k=0,\pm 1,\pm 2,\ldots\,$, and the  real parts $ \Re a_k $, $k=0,\pm 1,\pm 2,\ldots$ are such that
\[
 \ldots <\Re a_{k-1}<\Re a_k<\Re a_{k+1}<\ldots\,,\quad \lim_{k \to -\infty }\Re a_k=-\infty\,,\quad \lim_{k \to  \infty}\Re a_k= \infty\,.
\]
It is crucial that if the coefficient $h_k(A,M) $ or ${\widetilde h}_k(A,M) $ is independent of $A$, then 
\begin{eqnarray*}
&  &
\int_0^{ \widetilde\varepsilon  } h_k(A,M)\tau ^{ a_k}  \frac{\partial }{\partial r} (r\varphi  (r))\,dr=0 \quad \mbox{\rm or}\quad \int_0^{ \widetilde\varepsilon  } {\widetilde h}_k(A,M)\tau ^{ a_k}  \frac{\partial }{\partial r} (r\varphi  (r))\,dr=0 
\end{eqnarray*}
for every function $\varphi \in C_0^\infty(0,\widetilde\varepsilon) $. On the other hand, it will be shown that all coefficients $h_k(A,M) $, ${\widetilde h}_k(A,M) $ 
are polynomials in $\sqrt{1-A^2}$ or $\ln(1-A^2)$  (if $\ell <1$) or rational in $1-A^2$ (if $\ell >1$) functions. Thus, it is enough to prove  the existence of  the depending on $A$   coefficient  $h_k(A,M) $ or ${\widetilde h}_k(A,M) $. Then  the existence of  a function $\varphi \in C_0^\infty(0,\widetilde\varepsilon) $ such that  
\[ 
\Psi (0,t)
=  
\int_0^{ \widetilde\varepsilon  } \left(\frac{\partial }{\partial t}   K_1 \left(r,t; \pm m ;\varepsilon \right)
\right)  \frac{\partial }{\partial r} (r\varphi  (r))\,dr \not= 0 \quad \mbox{\rm for sufficiently large}\quad t\,,
\]  
is evident. This completes the proof of the necessity part of Theorem~\ref{T0.3}.

\section{Asymptotics of $\frac{\partial }{\partial t}   K_1 \left(r,t; \pm m ;\varepsilon \right) $ when $\ell >1$}
\label{S4}

In this section we  show that if $\ell >1$, then   the principal term of the  asymptotics of $\frac{\partial }{\partial t}   K_1 \left(r,t; \pm m ;\varepsilon \right) $
for the case of $M\not= 0,\pm 1$ is a function of  $r$. 
Taking into account the scaling,  henceforth we can  suppose $\varepsilon =1$.  
Denote
\begin{eqnarray}
\hspace{-0.9cm} & &
M:=\frac{i m}{\ell -1}, \quad A:=(\ell-1)r,\quad \tau  := t^{1-\ell}\to 0\quad \mbox{\rm as}\quad t \to \infty\,,
\nonumber \\
\hspace{-0.9cm} & &
\label{z12}
z:=
\frac{\left( \tau  -1\right)^2-A^2}{\left( \tau  +1\right)^2-A^2}=1-\frac{4 \tau }{1-A^2}+\frac{8 \tau ^2}{\left(1-A^2\right)^2}-\frac{4 \left(A^2+3\right) \tau ^3}{\left(1-A^2\right)^3}+O\left(\tau ^4\right)\quad \mbox{\rm as}\quad \tau  \to 0.
\end{eqnarray}
For the function defined by (\ref{K1def}) we obtain
\begin{eqnarray}
\label{K1der}
\frac{\partial }{\partial t}   K_1 \left(r,t; m ;1 \right)
& = &
C(m,\ell)t^{- \ell } \left( \left(1 +\tau \right)^2-A^2\right)^{M-2}{\mathcal F}(A,M;\tau ) , \quad C(m,\ell)\not=0\,,
\end{eqnarray}
where
\begin{eqnarray}
\label{Fcal}
{\mathcal F}(A,M;\tau )
& := &
2 M   \left(1-A^2- \tau ^2\right) F\left(1-M,1-M;2;\frac{\left( \tau  -1\right)^2-A^2}{\left( \tau  +1\right)^2-A^2}\right)\\
&  &
-\left(1 +\tau \right) \left(\left(1+\tau\right) ^2-A^2  \right) F\left(-M,-M;1;\frac{\left( \tau  -1\right)^2-A^2}{\left( \tau  +1\right)^2-A^2}\right)\,.\nonumber 
\end{eqnarray}
For the case of minus, $ M:=-\frac{i m}{\ell -1} $,  we have similar representation for the derivative
$
\frac{\partial }{\partial t}   K_1 \left(r,t; -m ;1 \right)$. 
We note, that for the values of $M=0,\pm 1 $, the function $\frac{\partial }{\partial t}   K_1 \left(r,t; \pm m ;1 \right)$ is independent of $r$:
 \begin{eqnarray}
 \label{17}
 & &
\frac{\partial}{\partial \tau } \left(\left((\tau +1)^2-A^2\right)^M F\left(-M,-M;1;\frac{(\tau -1)^2-A^2}{(\tau +1)^2-A^2}\right)\right) =
0 \quad if \quad  M=0\,,\\
 \label{18}
 & &
\frac{\partial}{\partial \tau } \left(\left((\tau +1)^2-A^2\right)^M F\left(-M,-M;1;\frac{(\tau -1)^2-A^2}{(\tau +1)^2-A^2}\right)\right) =
4 \tau \quad if \quad  M=1\,,\\
 \label{19}
 & &
\frac{\partial}{\partial \tau } \left(\left((\tau +1)^2-A^2\right)^M F\left(-M,-M;1;\frac{(\tau -1)^2-A^2}{(\tau +1)^2-A^2}\right)\right) =
-\frac{1}{4 \tau ^2} \quad if \quad  M=-1\,.
\end{eqnarray}
This indicates the fact that for these values of $M$ the Dirac equation obeys the Huygens' principle.

\subsection{The case of $\ell >1$ and $M \in {\mathbb C}$, $M\not=\frac{1}{2}k $, $k=0, \pm 1,\pm 2,\ldots$}

We note here that 
 \begin{eqnarray}
 \label{23less1}
 \left((\tau +1)^2-A^2\right)^M=\left(1-A^2\right)^M+2 M \tau  \left(1-A^2\right)^{M-1}-M \tau ^2 \left(1-A^2\right)^{M-2} \left(A^2-2 M+1\right)+O\left(\tau ^3\right)
\end{eqnarray}
as $\tau \searrow 0 $,  that together with (\ref{K1der}) allows us to ignore all terms of the expansion of the function ${\mathcal F}(A,M;\tau ) $ except the first one. 
\begin{proposition}
\label{P4.1}
Assume that $\ell >1$,   $M \in {\mathbb C}$,   $M\not=\frac{1}{2}k $, $k=0, \pm 1,\pm 2,\ldots$, and $A\in [0,1/2]$. The function ${\mathcal F}(A,M;\tau )$ (\ref{Fcal}) 
has the following expansion at $\tau =0 $:
\begin{eqnarray}
\label{19new}
{\mathcal F}(A,M;\tau )
&= &
\tau \frac{ 2 M  \Gamma (2 M)\left(A^2 (1-2 M)+1\right)}{(1-2 M) \Gamma (M+1)^2}\\
&  &
+\tau^{2 M} 16^M \frac{ \Gamma (1-2 M)}{\Gamma (1-M)^2}\left(1-A^2\right)^{-2 M} \left((A^2-1) +\tau   2(M-2)\right)
+O\left(\tau ^2\right) \left(\tau ^{2 M}+1\right)\,. \nonumber 
\end{eqnarray}
Then  
 the principal term in the series of the function ${\mathcal F}(A,M;\tau )$   is 
 \begin{eqnarray*}
\hspace{-0.3cm} &  & 
\tau^{2 M} 16^M 2 M\frac{ \Gamma (-2 M)}{\Gamma (1-M)^2}\left(1-A^2\right)^{1-2 M}  \quad if \quad \Re(M) <\frac{1}{2}\,, \\
\hspace{-0.3cm} &  &
\tau \frac{ 2 M  \Gamma (2 M)\left(A^2 (1-2 M)+1\right)}{(1-2 M) \Gamma (M+1)^2} \quad if \quad \Re(M) >\frac{1}{2} \,,\\
\hspace{-0.3cm} &  &
\tau\frac{(1+2 i B)   \left(2 A^2 B+i\right) \Gamma (1+2 i B)}{2 B \Gamma \left(\frac{3}{2}+i B\right)^2}
-\tau^{1+2 i B}\frac{ 2^{2+4 i B}   \Gamma (-2 i B)}{\Gamma \left(\frac{1}{2}-i B\right)^2}\left({1-A^2}\right)^{-2 i B}\,\, if \,\, M=\frac{1}{2}+iB,\,\, B \in \R\setminus \{0\}.  
\end{eqnarray*}
\end{proposition}
\medskip

\noindent
{\bf Proof.} In the function (\ref{Fcal})
we use the notation of $z$  from (\ref{z12}) and the expansion 
\begin{eqnarray*}
&  &
1-z= \frac{4\tau }{1-A^2}- \frac{8\tau^2 }{(1-A^2)^2}+O(\tau ^3)\,.
\end{eqnarray*}
The hypergeometric function is  defined (see, e.g., \cite[Sec.2.1]{B-E}) by the series
\begin{equation}
\label{defHGF}
F  \left( a,b;c;z  \right)= \sum_{n=0}^{\infty} \frac{ (a )_n (b )_n}{(c)_nn !} z^{n}\,.
\end{equation}
If $z$ is fixed and $|z| <1$, then $F  \left( a,b;c;z  \right)$ is entire analytic function of the parameters $a$, $b$,
and $c$ in the complex plane $\C$. (See, e.g., \cite[Sec.2.1.6]{B-E}.)
If $\Re (c-a-b)>0 $, then $F  \left( a,b;c;1 \right)=\frac{\Gamma (c)\Gamma (c-a-b)}{\Gamma (c-a)\Gamma (c-b)}$.

There is a 
formula (\cite[Sec.2.10]{B-E}) 
that ties together the values of $F  \left( a,b;c;z  \right) $ at the points $z=0$ and $z=1$:
\begin{eqnarray*} 
 F  \left( a,b;c;z  \right) 
& = &
\frac{\Gamma (c)\Gamma (c-a-b)}{\Gamma (c-a)\Gamma (c-b)}F  \left( a,b;a+b-c+1;1-z  \right) \\
 &  &
 + (1-z)^{c-a-b}\frac{\Gamma (c)\Gamma (a+b-c)}{\Gamma (a)\Gamma (b)}F  \left( c-a,c-b;c-a-b+1;1-z  \right) \,, \nonumber  
\end{eqnarray*}
where $|\arg (1-z)| <\pi$, $\quad |1-z|<1 $,  
and  $c-a-b \not= \pm 1,\pm 2,\ldots\,$. 
Hence, 
\begin{eqnarray}
\label{FminusM}
F\left(-M,-M;1;z\right)
& = &
\frac{\Gamma (1+2M)}{\Gamma (1+M)}F  \left( -M,-M;-2M;1-z  \right) + (1-z)^{1+2M}\frac{\Gamma (-2M-1)}{\Gamma (-M)^2 }\\
 &  &
\hspace{3cm}  \times F  \left( 1+M,1+M;2+2M;1-z  \right), \,\, 1+2M\not= \pm 1,\pm 2,\ldots\,,\nonumber \\
\label{F1minusM} 
F\left(1-M,1-M;2;z\right)
& = &
\frac{\Gamma (2M)}{\Gamma (M)^2}F  \left( 1-M,1-M;-2M+1;1-z  \right) \\
 &  &
 + (1-z)^{2M}\frac{\Gamma (-2M)}{\Gamma (1-M)^2}F  \left(1+M,1+M;2M+1;1-z  \right) , \quad 2M\not= \pm 1,\pm 2,\ldots \nonumber \,. 
\end{eqnarray}
It follows
\begin{eqnarray*}
&  &
F\left(-M,-M;1;\frac{\left( \tau  -1\right)^2-A^2}{\left( \tau  +1\right)^2-A^2}\right)\\
& = &
\frac{\Gamma (2 M+1)}{\Gamma (M+1)^2} \left(1-\frac{2 M  }{1-A^2}\tau\right)\\
&  &
-\left(1-A^2\right)^{-2 M-1} \tau ^{2 M+1}4^{2 M+1}\frac{ \Gamma (-2 M)}{(2 M+1) \Gamma (-M)^2}  \left(1-\frac{2 M  }{1-A^2}\tau\right)+\tau ^{2 M} O\left(\tau ^3\right)+O\left(\tau ^2\right)\,.
\end{eqnarray*}
Further,
\begin{eqnarray*}
&  &
F\left(1-M,1-M;2;\frac{\left( \tau  -1\right)^2-A^2}{\left( \tau  +1\right)^2-A^2}\right)\\
& = &
- \tau\frac{4  (M-1)^2 \Gamma (2 M)}{\left(A^2-1\right) (1-2 M) \Gamma (M+1)^2}+\frac{\Gamma (2 M)}{\Gamma (M+1)^2}+O\left(\tau ^2\right)\\
&  &
+\left(\frac{1-A^2}{\tau }\right)^{-2 M} \left(\frac{4^{2 M+1} \left(-M^2+M+1\right) \Gamma (1-2 M)}{(2 M+1) 2 \left(A^2-1\right) M \Gamma (1-M)^2} \tau -\frac{4^{2 M} \Gamma (1-2 M)}{2 M \Gamma (1-M)^2}+O\left(\tau ^2\right)\right)\,.
\end{eqnarray*}
Consequently,   (\ref{19new}) holds. 
Proposition is proved. 
\qed

\begin{lemma}
Assume that $\ell >1$,   $M \in {\mathbb C}$,   $M\not=\frac{1}{2}k $, $k=0, \pm 1,\pm 2,\ldots$, and $A\in [0,1/2]$. 
Then  
 the principal term in the asymptotic expansion  of the function $\frac{1}{C(m,\ell)}\frac{\partial }{\partial t}   K_1 \left(r,t; m ;1 \right)$   is 
 \begin{eqnarray*}
\hspace{-0.3cm} &  & 
t^{2 M(1-\ell)- \ell } 16^M 2 M\frac{ \Gamma (-2 M)}{\Gamma (1-M)^2}\left(1-A^2\right)^{-1-  M}  \quad if \quad \Re(M) <\frac{1}{2}\,, \\
\hspace{-0.3cm} &  &
t^{1- 2\ell }\frac{ 2 M  \Gamma (2 M)}{(1-2 M) \Gamma (M+1)^2}\left(A^2 (1-2 M)+1\right)\left(1-A^2\right)^{M-2} \quad if \quad \Re(M) >\frac{1}{2} \,,\\
\hspace{-0.3cm} &  &
t^{1- 2\ell }\frac{(1+2 i B)  \Gamma (1+2 i B)  }{2 B \Gamma \left(\frac{3}{2}+i B\right)^2}\left(2 A^2 B+i\right)\left(1-A^2\right)^{M-2}
-t^{(1+2 i B)(1-\ell)- \ell }\frac{ 2^{2+4 i B}   \Gamma (-2 i B)}{\Gamma \left(\frac{1}{2}-i B\right)^2}\left({1-A^2}\right)^{M-2-2 i B}\\
&  &
\hspace{8cm}  if \,\, M=\frac{1}{2}+iB,\,\, B \in \R\setminus \{0\}.  
\end{eqnarray*}
\end{lemma}
\medskip

\noindent
{\bf Proof.} It follows from    Proposition~\ref{P4.1} and (\ref{23less1}). \qed

\subsection{The case of $\ell >1$ and $M=n$, $n= 1 , 2,\ldots$}

In this case according to the  definition (\ref{K1def}) we have
\[
\frac{\partial}{ \partial t} K_1 \left(r,t; m ;\varepsilon \right)   
  =  
C(n,\ell,\varepsilon ) 
(1-\ell)t^{-\ell}\frac{\partial}{ \partial \tau }\left((\tau +1)^2-A^2\right)^n F\left(-n,-n;1;\frac{(\tau -1)^2-A^2}{(\tau +1)^2-A^2}\right)\,,\quad  C(n,\ell,\varepsilon )\not=0 .  
\] 
\begin{lemma}
For $n=1,2,\ldots$ the asymptotic expansion of the  function $\frac{\partial}{ \partial t} K_1 \left(r,t; m ;\varepsilon \right) $ contains the term  
\[
-C(n,\ell,\varepsilon )(1-\ell)t^{1-2\ell}  \frac{4 n^2 \Gamma (2 n) }{(2 n-1) \Gamma (n+1)^2}\left(1-A^2\right)^{n-2} \left(A^2 (2 n-1)-1\right)\,.
\]
\end{lemma}
\medskip

\noindent
{\bf Proof.} 
For $n=1,2,\ldots$ the hypergeometric function $F(-n,-n;1;z)$ is the following  polynomial 
\begin{equation} 
\label{Fpolyn}
F(-n,-n;1;z)=1+\sum _{j=1}^n  \left(\frac{\Gamma (n+1)}{\Gamma (j+1) \Gamma (n+1-j)}\right)^2z^j\,. 
\end{equation}
We use   (\ref{z12}) and obtain  
\begin{eqnarray*}
&  &
\frac{\partial}{ \partial t} K_1 \left(r,t; m ;\varepsilon \right)  \\
& = &
C(n,\ell,\varepsilon )(1-\ell)t^{-\ell}\frac{\partial }{\partial \tau }\Bigg[\left((\tau +1)^2-A^2\right)^n  \\
&  &
 \times  \left(1+\sum _{j=1}^n \left(1+\frac{4 \tau }{A^2-1}+\frac{8 \tau ^2}{\left(A^2-1\right)^2}+\frac{4 \left(A^2+3\right) \tau ^3}{\left(A^2-1\right)^3}+O(\tau ^4)\right)^j \left(\frac{\Gamma (n+1)}{\Gamma (j+1) \Gamma (n+1-j)}\right)^2\right)\Bigg]\\
& = &
-C(n,\ell,\varepsilon )(1-\ell)t^{-\ell} \left\{ \tau \frac{4 n^2 \left(1-A^2\right)^{n-2} \left(A^2 (2 n-1)-1\right) \Gamma (2 n)}{(2 n-1) \Gamma (n+1)^2}+O\left(\tau ^2\right)\right\} 
\end{eqnarray*}
for   $n=1,2,\ldots$. The lemma is proved. \qed

\subsection{The case of $\ell >1$ and $M=-n$, $n= 1 , 2,\ldots$}

If $M=-n$, $n=1,2,\ldots$, then according to the  definition (\ref{K1def}) we have
\[
 K_1 \left(r,t; m ;\varepsilon \right) 
 := 
C(n,\ell,\varepsilon ) 
\left((\tau +1)^2-A^2\right)^{-n} F\left( n, n;1;\frac{(\tau -1)^2-A^2}{(\tau +1)^2-A^2}\right)\,,\quad  C(n,\ell,\varepsilon )\not=0\,,
\] 
and
\begin{eqnarray}
\label{Mminusinteger}
&  &
\frac{\partial}{\partial \tau } \left(\left((\tau +1)^2-A^2\right)^{-n} F\left(n,n;1;\frac{(\tau -1)^2-A^2}{(\tau +1)^2-A^2}\right)\right)\\
& = &
-2 n (\tau +1) \left((\tau +1)^2-A^2\right)^{-n-1} F\left(n,n;1;\frac{(\tau -1)^2-A^2}{(\tau +1)^2-A^2}\right)\nonumber \\
&  &
+ 4 n^2 \left(A^2+\tau ^2-1\right) \left((\tau +1)^2-A^2\right)^{-n-2}F\left(n+1,n+1;2;\frac{(\tau -1)^2-A^2}{(\tau +1)^2-A^2}\right)\,.\nonumber 
\end{eqnarray}

\begin{lemma}
\label{L4.3}
The principal   term of the expansion of 
\begin{eqnarray*}
&  &
\frac{\partial}{\partial \tau } \left(\left((\tau +1)^2-A^2\right)^{-n} F\left(n,n;1;\frac{(\tau -1)^2-A^2}{(\tau +1)^2-A^2}\right)\right)
\end{eqnarray*}
as $\tau \to 0 $ is
\begin{eqnarray*} 
&  &
\cases{
\dsp -\frac{1}{4 }\tau ^{-2}, \quad \mbox{\rm if}\quad n=1\,,\cr
\dsp - 4^{1-2 n} n^2 \left(1-A^2\right)^{ n-1}
\frac{\Gamma (2n )}{[\Gamma (n+1)]^2}\tau  ^{-2 n}, \quad \mbox{\rm if}\quad  n=2,3,4\ldots \,.}
\end{eqnarray*}
\end{lemma}
\medskip

\noindent
{\bf Proof.} 
For $n=1$ we use (\ref{defHGF}) 
and  
$
F(1,1;1;z)=\frac{1}{1-z}
$. 
Hence
\[
\frac{\partial}{\partial \tau } \left(\left((\tau +1)^2-A^2\right)^{-1} F\left(1,1;1;\frac{(\tau -1)^2-A^2}{(\tau +1)^2-A^2}\right)\right)=
-\frac{1}{4 \tau ^2}\,.
\]
For $n=2$,
\[
F  \left( 2,2;1;z  \right)=  \sum_{n=0}^{\infty} \frac{\Gamma(2+n) \Gamma(2+n)}{\Gamma(1+n)} \frac{z^{n}}{n !}=  \sum_{n=0}^{\infty} (1+n)^2    z^{n}=\frac{z+1}{(1-z)^3}
\]
and, consequently, 
\begin{equation} 
\label{25}
\frac{\partial }{\partial \tau }\left(\left((\tau +1)^2-A^2\right)^{-2} F\left(2,2;1;\frac{(\tau -1)^2-A^2}{(\tau +1)^2-A^2}\right)\right) 
=-\frac{3 \left(1-A^2 \right)}{32 \tau ^4}-\frac{1}{32 \tau ^2}\,.
\end{equation}
Consider the case of  $n=3,4,\ldots \, $. There is a typo in the first statement of Lemma~6.3~\cite{JPHA2021}. The correct version is  
\begin{eqnarray*} 
  \lim_{x\searrow  0} x^{2 n-1} F (n,n;1;1-x)
& = &
 F (1-n,1-n;1;1)=\frac{\Gamma (2n-1)}{[\Gamma (n)]^2}, \quad n=2,3,\ldots\,,\\
  \lim_{x\searrow  0} x^{2 n-2} F (n,n;2;1-x)
& = &
 F (2-n,2-n;2;1)=\frac{\Gamma (2n-2)}{[\Gamma (n)]^2}, \quad n=2,3,\ldots\,.
\end{eqnarray*}
For the proof we appeal to \cite[(14) Sec. 2.1]{B-E} and to elementary relations, which can be verified, for instance, by (\ref{defHGF}) and the multiplication of the series in the left-hand  sides of the relations 
\begin{eqnarray}
\label{Finteger}
& & 
 (1-z)^{2 n-1} F(n,n;1;z)=F(1-n,1-n;1;z) \quad \mbox{\rm for all}\quad 0<z<1,\quad n=1,2,3,\ldots\,, \\
 &  &
(1-z)^{2 n-2} F(n,n;2;z)=F(2-n,2-n;2;z) \quad \mbox{\rm for all}\quad 0<z<1,\quad n=1,2,3,\ldots  \nonumber
\end{eqnarray}
or by \cite[(14) Sec. 2.10]{B-E}. Then with $a=b=n$, $c=1$,  it holds  $a+b-c=2n-1\geq 3 $
and we have 
\begin{eqnarray*}
F(n, n, 1 ; z)  
&= & 
(1-z)^{-2n+1}\frac{\Gamma(2n-1)}{[\Gamma(n)]^2 } \sum_{k=0}^{2n-2} \frac{[(-n+1)_{k}]^2 }{(-2n+2)_{k} k !}(1-z)^{k}    
 \,,  
\end{eqnarray*}
while with  $a=b=n$, $c=2$,  it holds  $a+b-c=2n-2\geq 3 $ and 
we have 
\begin{eqnarray*}
F(n, n, 2 ; z)  
&= & 
(1-z)^{-2n+2}\frac{\Gamma(2n-2)}{[\Gamma(n)]^2 } \sum_{k=0}^{2n-2} \frac{[(-n+2)_{k}]^2 }{(-2n+3)_{k} k !}(1-z)^{k}    
 \,, 
\end{eqnarray*}
where 
$-\pi<\arg (1-z)<\pi$.
According to (\ref{z12}) we can write 
\[
1-z=1-\frac{(\tau -1)^2-A^2}{(\tau +1)^2-A^2}, \quad \lim_{\tau \to 0} (1-z )=0\,.
\]
Thus, from (\ref{Mminusinteger}) we derive 
\begin{eqnarray*}
&  &
\frac{\partial}{\partial \tau } \left(\left((\tau +1)^2-A^2\right)^{-n} F\left(n,n;1;\frac{(\tau -1)^2-A^2}{(\tau +1)^2-A^2}\right)\right) \nonumber \\
& = &
-2 n (\tau +1) \left((\tau +1)^2-A^2\right)^{-n-1} \left(\frac{\Gamma (2n-1)}{[\Gamma (n)]^2} +o(1)\right)\left(1-\frac{(\tau -1)^2-A^2}{(\tau +1)^2-A^2}\right)^{1-2 n} \nonumber \\
&  &
+ 4 n^2 \left(A^2+\tau ^2-1\right) \left((\tau +1)^2-A^2\right)^{-n-2}\left(\frac{\Gamma (2n )}{[\Gamma (n+1)]^2}+o(1)\right)\left(1-\frac{(\tau -1)^2-A^2}{(\tau +1)^2-A^2}\right)^{-2 n}\,.
\end{eqnarray*}
Furthermore, there are easily verified, for instance, by induction, the following expansions
\begin{eqnarray*} 
\left(1-\frac{(\tau -1)^2-A^2}{(\tau +1)^2-A^2}\right)^{-2 n}
& = &
\left( \frac{\tau }{1-A^2}\right)^{-2 n} \left(4^{-2 n}-\frac{4^{1-2 n} n \tau }{A^2-1}+O\left(\tau ^2\right)\right),
\quad n>1,\\
\left(1-\frac{(\tau -1)^2-A^2}{(\tau +1)^2-A^2}\right)^{1-2 n}
& = &
\left(\frac{\tau }{1-A^2}\right)^{1-2 n} \left(4^{1-2 n}+\frac{2^{3-4 n} (1-2 n) \tau }{A^2-1}+O\left(\tau ^2\right)\right),
\quad n>2,\\
 \left((\tau +1)^2-A^2\right)^{-n-1}
 &=&
 \left(1-A^2\right)^{-n-1}-2 \tau  \left((n+1) \left(1-A^2\right)^{-n-2}\right)+O\left(\tau ^2\right),\\
  \left(A^2+\tau ^2-1\right) \left((\tau +1)^2-A^2\right)^{-n-2}
  & = &
 -\left(1-A^2\right)^{-n-1}+2 (n+2) \tau  \left(1-A^2\right)^{-n-2}+O\left(\tau ^2\right) .
\end{eqnarray*}
Hence,
\begin{eqnarray*}
&  &
\frac{\partial}{\partial \tau } \left(\left((\tau +1)^2-A^2\right)^{-n} F\left(n,n;1;\frac{(\tau -1)^2-A^2}{(\tau +1)^2-A^2}\right)\right)\\
& = &
-2 n (\tau +1) \Big[\left(1-A^2\right)^{-n-1}-2 \tau  (n+1) \left(1-A^2\right)^{-n-2}+O\left(\tau ^2\right) \Big]\\
&  &
\times  \left(\frac{\Gamma (2n-1)}{[\Gamma (n)]^2} +o(1)\right)\left( \frac{\tau }{1-A^2 }\right)^{1-2 n} \left(4^{1-2 n}+\frac{2^{3-4 n} (1-2 n) \tau }{A^2-1}+O\left(\tau ^2\right)\right)\\
&  &
+ 4 n^2 \Big[-\left(1-A^2\right)^{-n-1}+2 (n+2) \tau  \left(1-A^2\right)^{-n-2}+O\left(\tau ^2\right)\Big]\\
&  &
\times \left(\frac{\Gamma (2n )}{[\Gamma (n+1)]^2}+o(1)\right)\left( \frac{\tau }{1-A^2}\right)^{-2 n} \left(4^{-2 n}-\frac{4^{1-2 n} n \tau }{A^2-1}+O\left(\tau ^2\right)\right)\,.
\end{eqnarray*}
The principal term of the last asymptotics for $n=3,4,\ldots$ is 
\begin{eqnarray*}
&  &
- 4^{1-2 n} n^2 \left(1-A^2\right)^{ n-1}
\frac{\Gamma (2n )}{[\Gamma (n+1)]^2}\tau  ^{-2 n} \,.
\end{eqnarray*}
This completes the proof of the lemma.\qed

\subsection{The case of $\ell >1$ and $M=\frac{1}{2}+n $, $n=0, 1 , 2,\ldots$}

The   following lemma is decisive  for the asymptotics of the function $\frac{\partial}{ \partial t} K_1 \left(r,t; m ;\varepsilon \right)$.
\begin{lemma}
For  $A \in [0,1/2]$ and  all $n = 1,  2, 3,\ldots$ 
\begin{eqnarray*}
&  &
\frac{\partial}{\partial \tau } \left(\left((\tau +1)^2-A^2\right)^{\frac{1}{2}+n} F\left(-n-\frac{1}{2},-n-\frac{1}{2};1;\frac{(\tau -1)^2-A^2}{(\tau +1)^2-A^2}\right)\right)\\
& = &
- \tau (2 n+1)^2 \left(1-A^2\right)^{n-\frac{3}{2}} \left(2 A^2 n-1\right)\frac{ \Gamma (2 n)}{\Gamma \left(n+\frac{3}{2}\right)^2}+O\left(\tau ^2\right)\,.
\end{eqnarray*}
\end{lemma}
\smallskip

\noindent
{\bf Proof.} Indeed, 
\begin{eqnarray*}
&  &
\frac{\partial}{\partial \tau } \left(\left((\tau +1)^2-A^2\right)^{\frac{1}{2}+n} F\left(-n-\frac{1}{2},-n-\frac{1}{2};1;\frac{(\tau -1)^2-A^2}{(\tau +1)^2-A^2}\right)\right)\\
& = &
(2 n+1) (\tau +1) \left((\tau +1)^2-A^2\right)^{n-\frac{1}{2}} F\left(-n-\frac{1}{2},-n-\frac{1}{2};1;\frac{(\tau -1)^2-A^2}{(\tau +1)^2-A^2}\right)\\
&  &
+(2 n+1)^2 \left(A^2+\tau ^2-1\right) \left((\tau +1)^2-A^2\right)^{n-\frac{3}{2}} F\left(\frac{1}{2}-n,\frac{1}{2}-n;2;\frac{(\tau -1)^2-A^2}{(\tau +1)^2-A^2}\right)\,.
\end{eqnarray*}
According to Lemma~7.1~\cite{JPHA2021},  for $z\searrow 0 $ the following formulas   hold
\begin{eqnarray*}  
F \left(-n-\frac{1}{2},-n-\frac{1}{2};1;1-z\right) 
& = &
\frac{ \Gamma (2n+2)}{[\Gamma (\frac{3}{2}+n)]^2}-z \left(-n-\frac{1}{2}\right)^2  \frac{ \Gamma ( 2n+1)}{[\Gamma (\frac{3}{2}+n)]^2}+O(z^2),
\hspace{0.4cm}   n =0,  1,  2, 3,\ldots\,,\\ 
F \left(-n+\frac{1}{2},-n+\frac{1}{2};2;1-z\right)  
& = &
\frac{ \Gamma (2n+1)}{[\Gamma (\frac{3}{2}+n)]^2}
-z    \left(-n+\frac{1}{2}\right)^2   \frac{  \Gamma (2n)}{[\Gamma (\frac{3}{2}+n)]^2} +O(z^2) ,
\hspace{0.4cm}   n = 1,  2, 3,\ldots\,.
\end{eqnarray*}
Then
\begin{eqnarray*}
F\left(-n-\frac{1}{2},-n-\frac{1}{2};1;1-\frac{4 \tau }{(\tau +1)^2-A^2}\right)
& = &
\frac{\Gamma (2 n+2)}{\Gamma \left(n+\frac{3}{2}\right)^2}-\tau  \frac{\left( n+\frac{1}{2}\right)^2 4 \Gamma (2 n+1)}{\left((\tau +1)^2-A^2\right) \Gamma \left(n+\frac{3}{2}\right)^2}+O\left(\tau ^2\right),\\
F\left(\frac{1}{2}-n,\frac{1}{2}-n;2;1-\frac{4 \tau }{(\tau +1)^2-A^2}\right)
&  =&
\frac{\Gamma (2 n+1)}{\Gamma \left(n+\frac{3}{2}\right)^2}-\tau\frac{\left(\frac{1}{2}-n\right)^2 4   \Gamma (2 n)}{\left((\tau +1)^2-A^2\right) \Gamma \left(n+\frac{3}{2}\right)^2}+O\left(\tau ^2\right)\,.
\end{eqnarray*}
Hence,
\begin{eqnarray*}
&  &
(\tau +1) \left((\tau +1)^2-A^2\right)^{n-\frac{1}{2}} F\left(-n-\frac{1}{2},-n-\frac{1}{2};1;\frac{(\tau -1)^2-A^2}{(\tau +1)^2-A^2}\right)\\
&  &
+(2 n+1) \left(A^2+\tau ^2-1\right) \left((\tau +1)^2-A^2\right)^{n-\frac{3}{2}} F\left(\frac{1}{2}-n,\frac{1}{2}-n;2;\frac{(\tau -1)^2-A^2}{(\tau +1)^2-A^2}\right)\\
& = &
(\tau +1) \left((\tau +1)^2-A^2\right)^{n-\frac{1}{2}} \Bigg[\frac{\Gamma (2 n+2)}{\Gamma \left(n+\frac{3}{2}\right)^2}-\tau  \frac{\left( n+\frac{1}{2}\right)^2 4 \Gamma (2 n+1)}{\left((\tau +1)^2-A^2\right) \Gamma \left(n+\frac{3}{2}\right)^2}+O\left(\tau ^2\right) \Bigg]\\
&  &
+(2 n+1) \left(A^2+\tau ^2-1\right) \left((\tau +1)^2-A^2\right)^{n-\frac{3}{2}} \Bigg[\frac{\Gamma (2 n+1)}{\Gamma \left(n+\frac{3}{2}\right)^2}-\tau\frac{\left(\frac{1}{2}-n\right)^2 4   \Gamma (2 n)}{\left((\tau +1)^2-A^2\right) \Gamma \left(n+\frac{3}{2}\right)^2}+O\left(\tau ^2\right) \Bigg] \\
& = &
- \tau (2 n+1) \left(1-A^2\right)^{n-\frac{3}{2}} \left(2 A^2 n-1\right)\frac{ \Gamma (2 n)}{\Gamma \left(n+\frac{3}{2}\right)^2}+O\left(\tau ^2\right)\,.
\end{eqnarray*}
Lemma is proved.\qed

\subsection{The case of $\ell >1$ and $M=-\frac{1}{2}-n $, $n=0, 1 , 2,\ldots$}

For $n=1,2,\ldots$, from  (\ref{K1der}) we obtain the following relation
\[
\frac{\partial }{\partial t}   K_1 \left(r,t; m ;1 \right)
  =   
C(n,\ell,\varepsilon ) 
t^{-\ell} \left( \left(1 +\tau \right)^2-A^2\right)^{-\frac{5}{2}-n}{\mathcal F}\left(A,-\frac{1}{2}-n;\tau \right), \quad C(n,\ell,\varepsilon )\not=0\,.
\] 
\begin{lemma}
\label{L4.6}
There exist  numbers $C_0( \ell,\varepsilon )\not= 0 $ and $C_1(n,\ell,\varepsilon )\not=0 $  such that
if $n=0$, then
\[
\frac{\partial }{\partial t}   K_1 \left(r,t; m ;1 \right)
  =  
C_0( \ell,\varepsilon ) 
(1-A^2)^{-\frac{1}{2}}   t^{-1}+t^{-\ell} \ln (t) O\left( 1 \right),
\] 
while 
\[
\frac{\partial }{\partial t}   K_1 \left(r,t; m ;1 \right)
  =  
C_1(n,\ell,\varepsilon ) 
\left(1-A^2\right)^{-\frac{1}{2}-n }t^{-\ell}(1+o(1))\quad \mbox{  as}\quad t \to \infty\quad   \mbox{  if}\quad   n=1,2,\ldots  \,.   \nonumber  
\]
\end{lemma}
\medskip

\noindent
{\bf Proof.} First, we use (\ref{23less1}) : 
 \begin{eqnarray*}
&  &
\left( \left(1 +\tau \right)^2-A^2\right)^{-n-\frac{5}{2} } \\
& = &
\left(1-A^2\right)^{-n-\frac{5}{2}}-(2 n+5) \tau  \left(1-A^2\right)^{-n-\frac{7}{2}}+\frac{1}{2} (2 n+5) \tau ^2 \left(1-A^2\right)^{-n-\frac{9}{2}} \left(A^2+2 n+6\right)+O\left(\tau ^3\right) \,. 
\end{eqnarray*}
If we plug $M=-\frac{1}{2}-n $, $n=0, 1 , 2,\ldots$ in the definition (\ref{Fcal}) of ${\mathcal F}(A,M;\tau ) $,  
then 
\begin{eqnarray*}
{\mathcal F}\left(A,-\frac{1}{2}-n;\tau \right)
& = &
-(2n+1)   \left(1-A^2- \tau ^2\right) F\left(n+\frac{3}{2},n+\frac{3}{2};2;\frac{\left( \tau  -1\right)^2-A^2}{\left( \tau  +1\right)^2-A^2}\right)\\
&  &
-\left(1 +\tau \right) \left(\left(1+\tau\right) ^2-A^2  \right) F\left(n+\frac{1}{2},n+\frac{1}{2};1;\frac{\left( \tau  -1\right)^2-A^2}{\left( \tau  +1\right)^2-A^2}\right), \quad n=0 ,1, 2,\ldots
\end{eqnarray*}
Denote
\[
z:=\frac{\left( \tau  -1\right)^2-A^2}{\left( \tau  +1\right)^2-A^2} \in (0,1),\quad 1-z=\frac{4 \tau }{(\tau +1)^2-A^2}\to 0 \quad \mbox{\rm as}\quad \tau \to 0.
\] 
For $n=1,2,3,\ldots$, the following asymptotics
\begin{eqnarray*} 
&  &
\left(1+\tau  \right) \left(\left(1+\tau  \right)^2-A^2\right)  
F \left(n+\frac{1}{2},n+\frac{1}{2};1;\frac{\left(1-\tau \right)^2-A^2}{\left(1+\tau \right)^2-A^2}\right)\\
&  &
+(2n+1) \left(1-\tau ^{2}-A^2 \right)  
F \left(n+\frac{3}{2},n+\frac{3}{2};2;\frac{\left(1-\tau \right)^2-A^2}{\left(1+\tau \right)^2-A^2}\right) \\
& = &
(2n+1) \left(1-\tau ^{2}-A^2 \right)(1-z)^{- (2n+1)}\left\{ \frac{\Gamma(2n+1)}{[\Gamma(n+\frac{3}{2}) ]^2}
+ (1-z)  O(1)   \right\}  
\end{eqnarray*} 
 is proved in \cite[subsection~7.1, p.30]{JPHA2021}.
In particular,
\begin{eqnarray} 
\label{29}
&  &
\lim_{\tau  \to 0} (1-z)^{  (2n+1)}\Bigg[\left(1+\tau  \right) \left(\left(1+\tau  \right)^2-A^2\right)  
F \left(n+\frac{1}{2},n+\frac{1}{2};1;z\right)\\
&  &
+(2n+1) \left(1-\tau ^{2}-A^2 \right)  
F \left(n+\frac{3}{2},n+\frac{3}{2};2;z\right) \Bigg]  \nonumber \\
& = &
(2n+1) 
\frac{\Gamma(2n+1)}{[\Gamma(n+\frac{3}{2}) ]^2} \left(1-A^2 \right), \quad n=1,2,3,\ldots \nonumber  \,.
\end{eqnarray}
Together with
\begin{eqnarray*}
\left(1-\frac{(\tau -1)^2-A^2}{(\tau +1)^2-A^2}\right)^{-2 n-1}
& = &
\left( 1-A^2\right)^{2 n+1}\tau ^{-2 n-1}   4^{-2 n-1}  +\tau ^{-2 n}O\left(1\right) 
\end{eqnarray*}
the relation (\ref{29})  proves the second statement of Lemma~\ref{L4.6}.  
For the case of $n=0$ we apply (\ref{K1der}), (\ref{Fcal}), and the result of Section~8~\cite{JPHA2021}:
\begin{eqnarray*}
&  &
\left(1+\tau  \right) \left(\left(1+\tau  \right)^2-A^2\right)  
F \left(\frac{ 1}{2},\frac{ 1}{2};1;\frac{\left(1-\tau \right)^2-A^2}{\left(1+\tau \right)^2-A^2}\right)
+\left(1-\tau ^{2  }-A^2 \right)  
F \left(\frac{3}{2},\frac{3}{2};2;\frac{\left(1-\tau \right)^2-A^2}{\left(1+\tau \right)^2-A^2}\right)\\
& = &
\left(1+\tau  \right) \left(\left(1+\tau  \right)^2-A^2\right)  
  \left(\frac{1}{\pi}   
\left[ 2\psi (1) - 2\psi \left(\frac{ 1}{2}\right)  - \ln ( z) \right] +z\ln (z)O(1)\right)\\
&  &
+\left(1-\tau ^{2  }-A^2 \right)  
 \left(\frac{ 4z^{-1}}{ \pi }    
-O(1)\ln (z) \right)\\
& = &
\left(1-A^2 \right)^2  
    \frac{ 1}{ \pi }     \tau^{-1}
-O(1)\ln(t)\,.
\end{eqnarray*}
The lemma is proved. \qed

\subsection{The case of $\ell >1$ and $M=\frac{1}{2}$}

\begin{lemma}
For  $M=\frac{1}{2}$,
\begin{eqnarray*}
&  &
\frac{\partial}{\partial \tau } \left(((\tau +1)^2-A^2)^{\frac{1}{2}} F\left(-\frac{1}{2},-\frac{1}{2};1;\frac{(\tau -1)^2-A^2}{(\tau +1)^2-A^2}\right)\right)\\ 
& & 
=-4\frac{1}{\pi } \left(1-A^2\right)^{-3/2} \tau  \left(A^2-\ln  \left(4-4 A^2\right)+\ln  (\tau )+1\right)+O\left(\tau ^2\right)\,,
\end{eqnarray*}
and the principal  term of the expansion  is 
$
-\frac{4}{\pi } \left(1-A^2\right)^{-3/2} \tau \ln  (\tau )+O\left(\tau \right)\,.
$
\end{lemma}
\medskip

\noindent
{\bf Proof.} We apply the results of   subsection~7.3~\cite{JPHA2021} together with    (\ref{K1der}) and $M-2-3/2=-4$. \qed

\subsection{The case of $\ell >1$ and $M=-\frac{1}{2}$}

\begin{lemma}
The following asymptotic expansion holds:
\[
\frac{\partial }{\partial \tau }\left(\left((\tau +1)^2-A^2\right)^{-\frac{1}{2}} F\left(\frac{1}{2},\frac{1}{2};1;\frac{(\tau -1)^2-A^2}{(\tau +1)^2-A^2}\right)\right)
  =  
-\frac{1}{\pi  \sqrt{1-A^2} \tau }+O\left(\tau  \right)\quad \mbox{as} \quad \tau \to 0\,.
\]
\end{lemma}
\medskip

\noindent
{\bf Proof.} 
We have 
\begin{eqnarray}
\label{28new}
&  &
\frac{\partial }{\partial \tau }\left(\left((\tau +1)^2-A^2\right)^{-\frac{1}{2}} F\left(\frac{1}{2},\frac{1}{2};1;\frac{(\tau -1)^2-A^2}{(\tau +1)^2-A^2}\right)\right)\\
& = &
-\left((\tau +1)(\tau +1)^2-A^2\right)^{-\frac{3}{2}} F\left(\frac{1}{2},\frac{1}{2};1;\frac{(\tau -1)^2-A^2}{(\tau +1)^2-A^2}\right)\nonumber \\
&  &
+\left(A^2+\tau ^2-1\right) \left((\tau +1)^2-A^2\right)^{-\frac{5}{2}} F\left(\frac{3}{2},\frac{3}{2};2;\frac{(\tau -1)^2-A^2}{(\tau +1)^2-A^2}\right)\,. \nonumber 
\end{eqnarray}
For $ F\left(\frac{1}{2},\frac{1}{2};1;z\right)$  we apply   \cite[(12) Sec. 2.10]{B-E} with  $a=b=\frac{1}{2}$, $c=1$, $m=0$, and obtain
\begin{eqnarray}
\label{F12exp}
F\left(\frac{1}{2},\frac{1}{2};1;z\right)
& = &
\frac{1}{\pi } \sum_{n=0}^{\infty} \frac{[(\frac{1}{2})_{n}]^2}{[ n !]^2}\left[h_{n}^{\prime \prime}-\ln (1-z)\right](1-z)^{n} \\
& = &
\frac{4\ln  \left(2\right)-\ln  \left(1-z\right)}{\pi }+\frac{2-4 \ln  (2)+\ln  (1-z)}{4 \pi }(z-1)\nonumber\\
&  &
+\frac{3  (12\ln  (2)-7-3 \ln  (1-z))}{64 \pi }(z-1)^2+O\left((z-1)^3\right)\,. \nonumber
\end{eqnarray} 
If  $c-a-b=-m$,  then we apply \cite[(14) Sec. 2.10]{B-E}:
\begin{eqnarray*}
F(a, b, a+b-m ; z) \frac{1}{\Gamma(a+b-m)} 
&= & 
\frac{\Gamma(m)(1-z)^{-m}}{\Gamma(a) \Gamma(b)} \sum_{n=0}^{m-1} \frac{(a-m)_{n}(b-m)_{n}}{(1-m)_{n} n !}(1-z)^{n}    \\
&  &
+\frac{(-1)^ m}{\Gamma(a-m) \Gamma(b-m)} \sum_{n=0}^{\infty} \frac{(a)_{n}(b)_{n}}{(n+m)_{n} n !}\left[\bar{h}_{n}-\ln (1-z)\right](1-z)^{n} \,, \nonumber
\end{eqnarray*} 
where  
$-\pi<\arg (1-z)<\pi, \quad a, b, \neq 0,-1,-2, \dots
$, (\ref{barh}), 
and  $\sum_{n=0}^{m-1} $ is set zero if  $m=0$.

For the function $ F\left(\frac{3}{2},\frac{3}{2};2;z\right)$, we set $a=b=\frac{3}{2}$, $c=2$,  $m=1$, and obtain
\begin{eqnarray}
\label{F32exp}
F\left(\frac{3}{2},\frac{3}{2};2;z\right)
& = &
\frac{4}{\pi} (1-z)^{-1}       
-\frac{1}{\pi } \sum_{n=0}^{\infty} \frac{[(\frac{1}{2})_{n}]^2}{(n+1)_{n} n !}\left[\bar{h}_{n}-\ln (1-z)\right](1-z)^{n} \\
& = &
-\frac{4}{\pi  (z-1)}+\frac{3-2 \ln  (4)+\ln  (1-z)}{\pi }+\frac{3  (-17+12 \ln  (4)-6 \ln  (1-z))}{16 \pi }(z-1) \nonumber\\
&  &
-\frac{15  (-14+10 \ln  (4)-5 \ln  (1-z))}{64 \pi }(z-1)^2+O\left((z-1)^3\right)\,. \nonumber
\end{eqnarray}
Taking into account (\ref{z12}) we write
\begin{eqnarray*}
F\left(\frac{1}{2},\frac{1}{2};1;\frac{(\tau -1)^2-A^2}{(\tau +1)^2-A^2}\right)
& = &
\frac{\log \left(4 \left(1-A^2\right)\right)-\log (\tau )}{\pi }+\frac{  \left(-\log \left(4 \left(1-A^2\right)\right)+\log (\tau )+2\right)}{\pi  \left(A^2-1\right)}\tau+O\left(\tau ^2\right)\,,\\
F\left(\frac{3}{2},\frac{3}{2};2;\frac{(\tau -1)^2-A^2}{(\tau +1)^2-A^2}\right)
& = & 
\frac{1-A^2}{\pi  \tau }+\frac{5-\log \left(4 \left(1-A^2\right)\right)+\log (\tau )}{\pi }\,,\\
&  &
-\frac{ \left(4 A^2+18 \log \left(1-A^2\right)-18 \log (\tau )-47+36 \log (2)\right)}{4 \pi  \left(1-A^2\right)}\tau +O\left(\tau ^2\right)\,.
\end{eqnarray*} 
On the other hand, 

\begin{eqnarray*}
-(\tau +1) \left((\tau +1)^2-A^2\right)^{-\frac{3}{2}}
& = &
-\frac{1}{\left(1-A^2\right)^{3/2}}+\frac{\left(A^2+2\right)}{\left(1-A^2\right)^{5/2}} \tau +O\left(\tau ^2\right)\,,\\
\left(A^2+\tau ^2-1\right) \left((\tau +1)^2-A^2\right)^{-\frac{5}{2}}
& = &
-\frac{1}{\left(1-A^2\right)^{3/2}}+\frac{5  }{\left(1-A^2\right)^{5/2}}\tau+O\left(\tau ^2\right)\,.
\end{eqnarray*}
Then (\ref{28new}) imply
\begin{eqnarray*}
\frac{\partial }{\partial \tau }\left(\left((\tau +1)^2-A^2\right)^{-\frac{1}{2}} F\left(\frac{1}{2},\frac{1}{2};1;\frac{(\tau -1)^2-A^2}{(\tau +1)^2-A^2}\right)\right)
& = &
-\frac{1}{\pi  \sqrt{1-A^2} \tau }+O\left(\tau  \right)\,.
\end{eqnarray*}
The lemma is proved.\qed

\section{Asymptotics of $\frac{\partial }{\partial t}   K_1 \left(r,t; \pm m ;\varepsilon \right) $ when $\ell <1$}
\label{S5}

We appeal to the function $ K_1 \left(r,t; m ;\varepsilon \right)$ of  (\ref{K1def}). 
Taking into account the scaling,  henceforth we can  suppose $\varepsilon =1$.  
Assume $\ell <1$ and denote
\[ 
M:=\frac{i m}{\ell -1}, \quad A:=(\ell-1)r,\quad \tau  := t^{1-\ell}\to \infty\quad \mbox{\rm as}\quad t \to \infty,\quad 
z=\frac{\left( t^{1-\ell } -1\right)^2-A^2}{\left( t^{1-\ell } +1\right)^2-A^2}=\frac{\left( \tau  -1\right)^2-A^2}{\left( \tau  +1\right)^2-A^2} \,.
\]  
Then, according to (\ref{K1def}), we write
\[
 K_1 \left(r,t; m ;1 \right)  
  :=  
C(m,\ell)   \left( \left(1 +\tau \right)^2-A^2\right)^{M}   F \left(  -M,  -M;1;\frac{\left( \tau  -1\right)^2-A^2}{\left( \tau  +1\right)^2-A^2}\right), \quad C(m,\ell)\not=0  \,,  
\] 
and with the constant $C_1(m,\ell) =(1-\ell) C(m,\ell)$  obtain 
\[
\frac{\partial }{\partial t}   K_1 \left(r,t; m ;1 \right)
  =  
 C_1(m,\ell) t^{- \ell } \frac{\partial }{\partial \tau }\left( \left( \left(1 +\tau \right)^2-A^2\right)^{M}   F \left(  -M,  -M;1;\frac{\left( \tau  -1\right)^2-A^2}{\left( \tau  +1\right)^2-A^2}\right) \right) \nonumber \,.
\]
Further,  we consider
\[
 \frac{\partial }{\partial \tau }\left( \left( \left(1 +\tau \right)^2-A^2\right)^{M}   F \left(  -M,  -M;1;\frac{\left( \tau  -1\right)^2-A^2}{\left( \tau  +1\right)^2-A^2}\right) \right)
 = 
\left( \left(1 +\tau \right)^2-A^2\right)^{M-2}{\mathcal F}(A,M;\tau ) \nonumber \,,
\]
where
\begin{eqnarray*} 
{\mathcal F}(A,M;\tau )
& := &
2 M   \left(1-A^2- \tau ^2\right) F\left(1-M,1-M;2;\frac{\left( \tau  -1\right)^2-A^2}{\left( \tau  +1\right)^2-A^2}\right)\\
&  &
-\left(1 +\tau \right) \left(\left(1+\tau\right) ^2-A^2  \right) F\left(-M,-M;1;\frac{\left( \tau  -1\right)^2-A^2}{\left( \tau  +1\right)^2-A^2}\right)\,,
\end{eqnarray*}
similar to  (\ref{K1der}) and  (\ref{Fcal}), respectively.  We note  that any term of the asymptotics that is independent of $r$ obeys the Huygens' principle and can be neglected. (Compare with (\ref{17}),(\ref{18}), and  (\ref{19}).)

\subsection{The case of $\ell<1$ and $M \in {\mathbb C}$, $2M \not= 0,\pm 1,\pm 2,\ldots$}

\begin{proposition}
Assume that $\ell <1$ and $M \in {\mathbb C}$, $2M \not= 0,\pm 1,\pm 2,\ldots$, and  $A\in [0,1/3]$. Then   
  the following asymptotic expansion 
\begin{eqnarray}
\label{22}
&  &
 \frac{\partial }{\partial \tau }\left( \left( \left(1 +\tau \right)^2-A^2\right)^{M}   F \left(  -M,  -M;1;\frac{\left( \tau  -1\right)^2-A^2}{\left( \tau  +1\right)^2-A^2}\right) \right)\\
&=  &
\tau ^{2 M} \Bigg[-\frac{2 M \Gamma (1-2 M) \Gamma (2 M)}{  \Gamma (-2 M) [\Gamma (M+1)]^2}\tau^{-1} +\frac{4 (M-1) \left(A^2 (2 M-1)-1\right) \Gamma (2 M)}{(1-2 M)  [\Gamma (M)]^2}\tau ^{-3}+O\left( \frac{1}{\tau^5 }\right) \Bigg] \nonumber \\
&  &
-\frac{4^{2 M+1} \Gamma (1-2 M)}{2 (2 M+1) M  [\Gamma (-M)]^2}\tau ^{-2}-\frac{3 \left(4^{2 M+1} (M+1) \left(A^2 (2 M+3)+1\right) \Gamma (1-2 M)\right)}{2 (2 M+3) (2 M+1) M  [\Gamma (-M)]^2}\tau ^{-4}+O\left( \frac{1}{\tau^5 }\right)  \nonumber 
\end{eqnarray}
holds as $\tau \to \infty $.
\end{proposition}
\medskip

\noindent
{\bf Proof.} 
It is easily seen that
\begin{equation}
\label{321}
1-z
=
1-\frac{(\tau -1)^2-A^2}{(\tau +1)^2-A^2}=\frac{4}{\tau }-\frac{8}{\tau ^2}+\frac{4 \left(A^2+3\right)}{\tau ^3}-\frac{16 \left(A^2+1\right)}{\tau ^4}+\frac{4 \left(A^4+10 A^2+5\right)}{\tau ^5}+O\left(\frac{1}{\tau^6 }\right) 
\end{equation}
as $\tau \to \infty $.
We apply the previous expansion and the relations   (\ref{FminusM}), (\ref{F1minusM})  to obtain
\begin{eqnarray*}
&  &
F\left(-M,-M;1;\frac{(\tau -1)^2-A^2}{(\tau +1)^2-A^2}\right)\\
& = &
\frac{2 M \Gamma (2 M)}{\Gamma (M+1)^2}-\frac{4 M^2 \Gamma (2 M)}{ \Gamma (M+1)^2}\tau ^{-1}-\frac{8 M^2 \Gamma (2 M)}{(1-2 M)  \Gamma (M)^2}\tau ^{-2}\\
&  &
+\frac{4 M^2 \left(A^2 (6 M-3)+2 M (2 M (M+1)+1)-1\right) \Gamma (2 M)}{3 (1-2 M)  \Gamma (M+1)^2}\tau ^{-3}\\
&  &
-\frac{16 M^4 \left(A^2 (6 M-9)+M \left(M^2+M-1\right)-4\right) \Gamma (2 M)}{3 (2 M-3) (1-2 M)  \Gamma (M+1)^2}\tau ^{-4}+O\left(\frac{1}{\tau^5 }\right)\\
&  &
+\tau ^{-2 M} \Bigg[\frac{4^{2 M+1} \Gamma (1-2 M)}{2 M (2 M+1)  \Gamma (-M)^2} \tau^{-1}-\frac{2^{4 M+2} \Gamma (1-2 M)}{(2 M+1)  \Gamma (-M)^2}\tau ^{-2}\\
&  &
+\frac{4^{2 M+1} \left(A^2 (4 M (M+2)+3)+4 M (M+1)^2+1\right) \Gamma (1-2 M)}{2 (2 M+3) M (2 M+1)  \Gamma (-M)^2}\tau ^{-3}\\
&  &
-\frac{16^{M+1} (M+1) \left(A^2 (6 M+9)+2 M (M+2)+3\right) \Gamma (1-2 M)}{6 (2 M+3) (2 M+1)  \Gamma (-M)^2}\tau ^{-4}\\
&  &
+\frac{4^{2 M+1} (M+1) \Gamma (1-2 M)}{6 (2 M+3) (2 M+5) M (2 M+1)  \Gamma (-M)^2}\\
&  &
\times \Big(3 A^4 (2 M+1) (2 M+3) (2 M+5)+6 A^2 (2 M+3) (2 M+5) (2 M (M+1)+1)\\
&  &
+2 M (2 M (M+3)+3) (2 M (M+3)+7)+9 \Big)\tau ^{-5}+O\left(\frac{1}{\tau^6 }\right)\Bigg]
\end{eqnarray*}
and
\begin{eqnarray*}
&  &
F\left(1-M,1-M;2;\frac{(1-\tau )^2-A^2}{(\tau +1)^2-A^2}\right)\\
& = &
\frac{\Gamma (2 M)}{\Gamma (M+1)^2}+\frac{4 (M-1)^2 \Gamma (2 M)}{ \left((1-2 M) \Gamma (M+1)^2\right)} \tau^{-1} -\frac{4 (M-1) ((M-2) M+2) \Gamma (2 M)}{ \left((1-2 M) \Gamma (M+1)^2\right)} \tau ^{-2}\\
&  &
+\frac{4 (M-1) \left(3 A^2 (M-1) (2 M-3)+M (2 M (2 (M-4) M+17)-45)+27\right) \Gamma (2 M)}{3 (2 M-3) (1-2 M)  \Gamma (M+1)^2}\tau ^{-3}\\
&  &
-\frac{8(M-1)\Gamma (2 M)}{3  \left((2 M-3) (1-2 M) \Gamma (M+1)^2\right)} \\
&  &
\hspace{2.5cm}\times   \left(3 A^2 (2 M-3) ((M-2) M+2)+M (M (M ((M-4) M+11)-23)+30)-18\right)\tau ^{-4}\\
&  &
+O\left(\frac{1}{\tau^5 }\right)
+\tau ^{-2 M} \Bigg[-\frac{16^M \Gamma (1-2 M)}{2 M \Gamma (1-M)^2}+\frac{4^{2 M+1} ((M-1) M-1) \Gamma (1-2 M)}{2 M (2 M+1)   \Gamma (1-M)^2}\tau^{-1}\\
&  &
-\frac{2^{4 M} \left(M \left(A^2 (2 M+1)+2 (M-2) M+1\right)+4\right) \Gamma (1-2 M)}{M (2 M+1)  \Gamma (1-M)^2}\tau ^{-2}\\
&  &
+\frac{4^{2 M+1}\Gamma (1-2 M)}{6 (2 M+3) M (2 M+1) \Gamma (1-M)^2} \\
&  &
\hspace{2.5cm} \times \left(3 A^2 (2 M+1) (2 M+3) ((M-1) M-1)+(M-1) M \left(4 M^3-4 M+15\right)-27\right) \tau ^{-3} \\
&  &
-\frac{16^M \Gamma (1-2 M)}{6 (2 M+3) M (2 M+1) \Gamma (1-M)^2}\\
&  &
\times \Big(3 A^4 M (2 M+1)^2 (2 M+3)+6 A^2 (2 M+3) (M (4 M ((M-1) M-1)+11)+8)\\
&  &
+M (2 M (2 M (2 M ((M-1) M+1)+5)-47)+57)+144\Big)\tau ^{-4}  +O\left(\frac{1}{\tau^5 }\right)\Bigg]
\end{eqnarray*}
as $\tau \to \infty $. On the other hand,
\begin{eqnarray*}
\left( \left(1 +\tau \right)^2-A^2\right)^{M-2}
& = & \tau ^{2M- 4}\Bigg[ 1+\frac{2 (M-2)}{\tau }-\frac{(M-2) \left(A^2-2 M+5\right)}{\tau ^2}\\
&  &
-\frac{2 \left((M-3) (M-2) \left(3 A^2-2 M+5\right)\right)}{3 \tau ^3}\\
&  &
+\frac{(M-3) (M-2) \left(3 A^4-12 \left(A^2+2\right) M+42 A^2+4 M^2+35\right)}{6 \tau ^4}+O\left(\frac{1}{\tau^5 }\right)\Bigg]
\end{eqnarray*}
as $\tau \to \infty $. Then we put the  last expressions  into (\ref{Fcal}) and obtain (\ref{22}). 
The proposition is proved. \qed
\begin{corollary}
If $ \Re(M) >-\frac{1}{2}$, then in the expansion (\ref{22}) there is 
the term  containing $A$: 
\[
\tau ^{2 M-3}  \frac{4 (M-1) \left(A^2 (2 M-1)-1\right) \Gamma (2 M)}{(1-2 M)  [\Gamma (M)]^2} \,.
\]
If $ \Re(M) <-\frac{1}{2}$, then in the expansion  (\ref{22}) there is
the term containing $A$:
\[
-\frac{3 \left(4^{2 M+1} (M+1) \left(A^2 (2 M+3)+1\right) \Gamma (1-2 M)\right)}{2 (2 M+3) (2 M+1) M  [\Gamma (-M)]^2}\tau ^{-4}\,.
\]
If $  M=-\frac{1}{2}+iB ,\,\,  B\in \R\setminus \{0\}$, then in the expansion  (\ref{22}) there are
the terms  containing $A$:
\[
-\frac{1}{4 \tau ^4}\Bigg[ \frac{8 (2 B+3 i) (2 A^2 (B+i)+i) }{[\Gamma (-\frac{1}{2}+i B)]^2 \Gamma (3-2 i B)}\tau ^{2 i B}+\frac{3 i 4^{2 i B} (2 B-i) (2 A^2 (B-i)-i)}{(B-i) [\Gamma \left(\frac{1}{2}-i B\right)]^2 \Gamma (1+2 i B)}\Bigg]\,.
\]
\end{corollary}

\subsection{The case of   $\ell <1$ and $M=n$, $ n=2,3,\ldots$}

\begin{lemma}
The function
\begin{equation}
\label{29Kdernew}
\frac{\partial }{\partial \tau }\left( \left( \left(1 +\tau \right)^2-A^2\right)^{n}   F \left(  -n,  -n;1;\frac{\left( \tau  -1\right)^2-A^2}{\left( \tau  +1\right)^2-A^2}\right) \right) 
\end{equation}
for $n=2,3,\ldots$   is a polynomial in $\tau$ and $A^2$, which varies in $A^2$ and contains at least one term  $ \tau ^aA^{2b}$ with $a,b\not=0 $.   
\end{lemma}
\medskip

\noindent
{\bf Proof.} To verify the first statement we just use (\ref{Fpolyn}).
For  $M=n$, where $n=2,3,4,\ldots$ we  put $
z=\frac{\left( \tau  -1\right)^2-A^2}{\left( \tau  +1\right)^2-A^2}$ 
in  (\ref{Fpolyn}), and obtain for the function of (\ref{29Kdernew}) the polynomial
\begin{eqnarray*}
 &  &
 \frac{\partial }{\partial \tau }\left( \left( \left(1 +\tau \right)^2-A^2\right)^{n}\Bigg[ 1+\sum _{j=1}^n  \left(\frac{\Gamma (n+1)}{\Gamma (j+1) \Gamma (-j+n+1)}\right)^2\left( \frac{\left( \tau  -1\right)^2-A^2}{\left( \tau  +1\right)^2-A^2}\right)^j\Bigg]    \right)\\
& = &
2n(\tau+1) \left( \left(1 +\tau \right)^2-A^2\right)^{n-1}   \\
&  &
+ 2\sum _{j=1}^n  \left(\frac{n!}{ j!   (n-j)!}\right)^2 \Bigg[j(\tau-1)\left( \left( \tau  -1\right)^2-A^2\right)^{j-1}\left( \left(\tau+1 \right)^2-A^2\right)^{n-j}\\
&   &
+(n-j)(\tau+1)\left( \left( \tau  -1\right)^2-A^2\right)^j\left( \left(\tau+1 \right)^2-A^2\right)^{n-j-1}\Bigg]\,.  
\end{eqnarray*} 
Thus, the result is a polynomial in $\tau$ and $A^2$.  The coefficient of $A^{2n-1}$ vanishes since
\[
n+\sum _{j=1}^n (n-2 j) \left(\frac{\Gamma (n+1)}{\Gamma (j+1) \Gamma (-j+n+1)}\right)^2=0\,.
\]
At   $A=\tau-1 $ 
the polynomial equals to $[(2n +2)\tau + (2n -2)]\left( 4\tau\right)^{n-1} $, 
while at $ A=\tau +1$ it takes value $
(2n\tau-2n) \left( 4\tau\right)^{n-1}$. Thus, there is at least one nonvanishing term of the plynomial containing $ \tau ^aA^{2b}$ with $a,b\not=0 $. We can use such term with the greatest $a$, in fact $a=2n-3$.
The lemma is proved. \qed

\subsection{The case of   $\ell <1$ and $M=-n$, $ n= 1,2,3,\ldots$}

 In the case of $M=-n$, with $ n=1,2,3,\ldots$, we appeal to (\ref{Mminusinteger}) to prove the next lemma.

\begin{lemma} 
For $n=2,3,4,\ldots$,   in the expansion of the function 
\begin{equation}
\label{29Kder}
\frac{\partial}{\partial \tau } \left(\left((\tau +1)^2-A^2\right)^{-n} F\left(n,n;1;\frac{(\tau -1)^2-A^2}{(\tau +1)^2-A^2}\right)\right)
\end{equation}
as $\tau \to \infty $ there exists the term   
\begin{eqnarray*}
&  &
- \frac{2^{1-2 n} \left(1-A^2\right)^{n-1} \Gamma \left(n+\frac{1}{2}\right)}{\sqrt{\pi } \Gamma (n)}\frac{1}{\tau ^{2 n}}\,.
\end{eqnarray*} 
\end{lemma}
\medskip

\noindent
{\bf Proof.} The case of $M=-1, -2 \,$ is given by Lemma~\ref{L4.3} and (\ref{25}).
For the case of  $n=3,4,\ldots $ we appeal to \cite[(14) section 2.1]{B-E} and to elementary relation  (\ref{Finteger}).
Hence, by definition (\ref{defHGF}) we have
\[
  F(n,n;1;z)
  =  (1-z)^{1-2 n }F(1-n,1-n;1;z)=(1-z)^{1-2 n }\sum _{k=0}^{n-1} \frac{ [(1-n)_k]^2}{[k!]^2}z^k\,.
\]
Then
\begin{eqnarray*}
\frac{\partial}{\partial \tau } \left(\left((\tau +1)^2-A^2\right)^{-n} F\left(n,n;1;\frac{(\tau -1)^2-A^2}{(\tau +1)^2-A^2}\right)\right)
& = &
4^{1-2 n} \tau ^{-2 n}P(A;n,\tau) \,,
\end{eqnarray*}
where
\begin{eqnarray*}
P(A;n,\tau)& := &
 \tau \sum _{k=0}^{n-1} \frac{\left[(1-n)_k\right]^2}{[k! ]^2}  \left(A^2 ((n-1) (\tau +1)-2 k)-\left(\tau ^2-1\right) (2 k+(n-1) (\tau -1))\right)\\
&  &
+4^{1-2 n} (1-2 n) \sum _{k=0}^{n-1}\frac{\left[(1-n)_k\right]^2}{[k! ]^2}  \left(\left((\tau -1)^2-A^2\right)^k \left((\tau +1)^2-A^2\right)^{-k+n-1}\right) 
\end{eqnarray*}
is   polynomial in $\tau$ and $A^2$.
We set $\tau=0$ and obtain 
\[
P(A;n,0) 
  =  
-\frac{2^{1-2 n} \left(1-A^2\right)^{n-1} \Gamma \left(n+\frac{1}{2}\right)}{\sqrt{\pi } \Gamma (n)}\,.
\]
Hence, the asymptotic expansion of the function   (\ref{29Kder}) contains the term 
\[
-\tau ^{-2 n} \frac{2^{1-2 n} \left(1-A^2\right)^{n-1} \Gamma \left(n+\frac{1}{2}\right)}{\sqrt{\pi } \Gamma (n)}\,.
\]
The lemma is proved. \qed

Thus, in the asymptotic expansion of (\ref{29Kder}) we can choose the  term with greatest degree in $\tau$ among all containing  variable $A^2$.

\subsection{The case of $\ell <1$ and $M=\frac{1}{2}+n $, $n= 1 , 2,\ldots$}

\begin{lemma}
For $M= \frac{3}{2}$ ($n=1$) we have  
\[
\frac{\partial}{\partial \tau } \left(\left((\tau +1)^2-A^2\right)^{ -\frac{3}{2}} F\left(\frac{3}{2},\frac{3}{2};1;\frac{(\tau -1)^2-A^2}{(\tau +1)^2-A^2}\right)\right)\\
=
-\frac{1}{4 \pi  \tau ^2}+\frac{6 A^2-6 \ln  (\tau )+5-6 \ln  (4)}{16 \pi  \tau ^4}+O\left(\frac{1}{\tau ^5}\right).
\]
For $n=2,3,4,\ldots\,$,   in the asymptotic expansion of the function 
\begin{eqnarray*}
&  &
\frac{\partial}{\partial \tau } \left(\left((\tau +1)^2-A^2\right)^{-n-\frac{1}{2}} F\left(n+\frac{1}{2},n+\frac{1}{2};1;\frac{(\tau -1)^2-A^2}{(\tau +1)^2-A^2}\right)\right)
\end{eqnarray*}
as $\tau \to \infty $ there exists the term with $\tau ^{-4} $ depending on $A$, namely, this term is
\[
\frac{3\Gamma(2n)  \left(4^{-2 n-1} (2 n-1) \left(2 A^2 (n-1)-1\right)\right)}{[\Gamma(n+\frac{1}{2})]^2(n-1)}\frac{1}{ \tau^4 }\,.
\]
\end{lemma}
\medskip

\noindent
{\bf Proof.} 
For $n=1,2,3,4,\ldots$, we have to discuss
\[
\frac{\partial }{\partial \tau }\left(\left((\tau +1)^2-A^2\right)^{-n-\frac{1}{2}} F\left(n+\frac{1}{2},n+\frac{1}{2};1;\frac{(\tau -1)^2-A^2}{(\tau +1)^2-A^2}\right)\right)\,.
\]
We make the change of variable $\lambda =\tau ^{-1}$ and consider the  asymptotic series about $\lambda =0$:  
\begin{eqnarray*}
& &
\frac{\partial }{\partial \lambda }\left(\left(\left(\frac{1}{\lambda }+1\right)^2-A^2\right)^{-n-\frac{1}{2}} 
F\left( n+\frac{1}{2}, n+\frac{1}{2};1;\frac{\left(\frac{1}{\lambda }-1\right)^2-A^2}{\left(\frac{1}{\lambda }+1\right)^2-A^2}\right)\right)\\
&  = &
(2 n+1)\lambda ^{2 n}\left(\lambda  \left(-A^2\lambda+\lambda +2\right)+1\right)^{-n-\frac{5}{2}} {\mathcal F}(A,n;\lambda )\,,
\end{eqnarray*}
where has been used the notation 
\begin{eqnarray*}
{\mathcal F}(A,n;\lambda  )
& := &(\lambda +1) \left(\lambda  \left(-A^2\lambda+\lambda +2\right)+1\right) F\left(n+\frac{1}{2},n+\frac{1}{2};1;\frac{\left(\frac{1}{\lambda }-1\right)^2-A^2}{\left(\frac{1}{\lambda }+1\right)^2-A^2}\right)\\
&  &
-\lambda  ((2 n+1)) \left(\left(A^2-1\right) \lambda ^2+1\right) F\left(n+\frac{3}{2},n+\frac{3}{2};2;\frac{\left(\frac{1}{\lambda }-1\right)^2-A^2}{\left(\frac{1}{\lambda }+1\right)^2-A^2}\right)\,. \nonumber
\end{eqnarray*} 
Denote
\[
z:=\frac{\left(\frac{1}{\lambda }-1\right)^2-A^2}{\left(\frac{1}{\lambda }+1\right)^2-A^2} \to 1 \quad \mbox{as}\quad \lambda \to 0\,,
\]
then we can use (\ref{321}) as $\lambda =\tau^{-1} \to 0 $. 
According to  \cite[(14) Sec. 2.10]{B-E} for $n=1,2,\ldots$ with $a=b=n+\frac{1}{2}$, $c=1$, $c-a-b=-2n $, and $m=2n$,  we have :
\begin{eqnarray*} 
F\left(n+\frac{1}{2},n+\frac{1}{2};1;z\right)
&= & 
(1-z)^{-2n}\frac{\Gamma(2n)}{[\Gamma(n+\frac{1}{2})]^2} \sum_{k=0}^{2n-1} \frac{[(\frac{1}{2}-n)_{k}]^2}{(1-2n)_{k} k !}(1-z)^{k}    \\
&  &
+\frac{1}{[\Gamma(\frac{1}{2}-n)]^2} \sum_{k=0}^{\infty} \frac{[(n+\frac{1}{2})_{k}]^2}{(k+2n)_{k} k !}\left[\bar{h}_{k}-\ln (1-z)\right](1-z)^{k} \\
&= & 
(1-z)^{-2n}\frac{\Gamma(2n)}{[\Gamma(n+\frac{1}{2})]^2} \sum_{k=0}^{2n-1} \frac{[(\frac{1}{2}-n)_{k}]^2}{(1-2n)_{k} k !}(1-z)^{k}  
+|\ln (1-z)|O(1)\,, \nonumber
\end{eqnarray*} 
where  
$-\pi<\arg (1-z)<\pi, \quad a, b, \neq 0,-1,-2, \dots
$
\begin{equation}
\label{barh}
\bar{h}_{n}=\psi(1+n)+\psi(1+n+m)-\psi(a+n)-\psi(b+n)\,.
\end{equation}     
The function $\psi(z)$ is the logarithmic derivative of the gamma function: $
 \psi(z)=\frac{\Gamma^{\prime}(z)}{\Gamma(z)}$.

Similarly, 
  since  $a=b=n+\frac{3}{2}$, $c=2$, $c-a-b= -m$ , $m=2n+1$, we have
\begin{eqnarray*}
 F \left(n+\frac{3}{2},n+\frac{3}{2};2;z\right) 
&= & 
\frac{\Gamma(2n+1)(1-z)^{-2n-1}}{[\Gamma(n+\frac{3}{2})]^2 } \sum_{k=0}^{2n} \frac{[( \frac{1}{2}- n)_{k}]^2}{( -2n)_{k} k !}(1-z)^{k}    \\
&  &
+\frac{(-1)}{[\Gamma( \frac{1}{2}-n) ]^2} \sum_{k=0}^{\infty} \frac{[(n+\frac{3}{2})_{k}]^2}{(k+2n+1)_{k} k !}\left[\bar{h}_{k}-\ln (1-z)\right](1-z)^{k} \\
&= & 
(1-z)^{-2n-1}\frac{\Gamma(2n+1)}{[\Gamma(n+\frac{3}{2})]^2 } \sum_{k=0}^{2n} \frac{[( \frac{1}{2}- n)_{k}]^2}{( -2n)_{k} k !}(1-z)^{k}   
+|\ln (1-z)|O(1) \,. 
\end{eqnarray*} 
Hence,
\begin{eqnarray*}
&  &
{\mathcal F}(A,n;\lambda  )\\
&= &
 (1-z)^{-2n-1}\Bigg[ (\lambda +1) \left(\lambda  \left(-A^2\lambda+\lambda +2\right)+1\right) \\
&  &
\times  \Bigg\{\frac{\Gamma(2n)}{[\Gamma(n+\frac{1}{2})]^2} \sum_{k=0}^{2n-1} \frac{[(\frac{1}{2}-n)_{k}]^2}{(1-2n)_{k} k !}(1-z)^{k+1}  
+(1-z)^{2n+1}|\ln (1-z)|O(1)\Bigg\} \\
&  &
-\lambda  ((2 n+1)) \left(\left(A^2-1\right) \lambda ^2+1\right) \Bigg\{\frac{\Gamma(2n+1)}{[\Gamma(n+\frac{3}{2})]^2 } \sum_{k=0}^{2n} \frac{[( \frac{1}{2}- n)_{k}]^2}{( -2n)_{k} k !}(1-z)^{k}   
+(1-z)^{2n+1}|\ln (1-z)|O(1)  \Bigg\}\Bigg]\\
&= &
 (1-z)^{-2n-1}\frac{\Gamma(2n)}{[\Gamma(n+\frac{1}{2})]^2} \\
&  &
\times \Bigg[(\lambda +1) \left(\lambda  \left(-A^2\lambda+\lambda +2\right)+1\right)  \Bigg\{ \sum_{k=0}^{2n-1} \frac{[(\frac{1}{2}-n)_{k}]^2}{(1-2n)_{k} k !}(1-z)^{k+1}  
+(1-z)^{2n+1}|\ln (1-z)|O(1)\Bigg\} \\
&  &
-\lambda  ((2 n+1)) \left(\left(A^2-1\right) \lambda ^2+1\right) \Bigg\{\frac{2n}{[(n+\frac{1}{2})]^2} \sum_{k=0}^{2n} \frac{[( \frac{1}{2}- n)_{k}]^2}{( -2n)_{k} k !}(1-z)^{k}   
+(1-z)^{ 2n+1}|\ln (1-z)|O(1)  \Bigg\}\Bigg] \,. \nonumber
\end{eqnarray*} 
Since $n \geq 2$, we can truncate the sums as follows:
\begin{eqnarray*}
{\mathcal F}(A,n;\lambda  )
&= &
 (1-z)^{-2n-1}\frac{\Gamma(2n)}{[\Gamma(n+\frac{1}{2})]^2} \\
&  &
\times \Bigg[(\lambda +1) \left(\lambda  \left(-A^2\lambda+\lambda +2\right)+1\right)  \Bigg\{ \sum_{k=0}^{2} \frac{[(\frac{1}{2}-n)_{k}]^2}{(1-2n)_{k} k !}(1-z)^{k+1}  
+(1-z)^{4}|O(1)\Bigg\} \\
&  &
-\lambda  ((2 n+1)) \left(\left(A^2-1\right) \lambda ^2+1\right) \Bigg\{\frac{2n}{[(n+\frac{1}{2})]^2} \sum_{k=0}^{3} \frac{[( \frac{1}{2}- n)_{k}]^2}{( -2n)_{k} k !}(1-z)^{k}   
+(1-z)^{ 4}O(1)  \Bigg\}\Bigg] \,.
\end{eqnarray*} 
It is evident that for  $n \geq 2$ the following asymptotic expansion holds
\begin{eqnarray*}
&  &
(2 n+1)\lambda ^{2 n}\left(\lambda  \left( -A^2\lambda +\lambda +2\right)+1\right)^{-n-\frac{5}{2}}\left(1-\frac{\left(\frac{1}{\lambda }-1\right)^2-A^2}{\left(\frac{1}{\lambda }+1\right)^2-A^2}\right)^{-2 n-1}\\
& = &
4^{-2 n-1} (2 n+1)\frac{1}{\lambda }+4^{-2 n-1} (2 n-3) (2 n+1)+  2^{-4 n-3} (2 n-3) (2 n+1) \left(-A^2+2 n-4\right)\lambda+O\left(\lambda ^2\right)\,.
\end{eqnarray*} 
On the other hand
\begin{eqnarray*}
&  &
(\lambda +1) \left(\lambda  \left( -A^2\lambda+\lambda +2\right)+1\right)  \Bigg\{ \sum_{k=0}^{2} \frac{[(\frac{1}{2}-n)_{k}]^2}{(1-2n)_{k} k !}(1-z)^{k+1}  
+(1-z)^{4}O(1)\Bigg\} \\
&  &
-\lambda  ((2 n+1)) \left(\left(A^2-1\right) \lambda ^2+1\right) \Bigg\{\frac{2n}{[(n+\frac{1}{2})]^2} \sum_{k=0}^{3} \frac{[( \frac{1}{2}- n)_{k}]^2}{( -2n)_{k} k !}(1-z)^{k}   
+(1-z)^{4}O(1)  \Bigg\}\\
& = & 
\frac{\lambda  (\lambda +1)}{(n-1) \left(\lambda  \left(A^2 (-\lambda )+\lambda +2\right)+1\right)^2}\\
&  &
\times \Bigg[ \Big(\lambda ^2 \left(-8 A^2 (n-1)+8 n^3-44 n^2+78 n-41\right)\\
&  &
+4 \left(A^2-1\right)^2 \lambda ^4 (n-1)+4 \left(A^2-1\right) \lambda ^3 (n-1) (2 n-5)-4 \lambda  (n-1) (2 n-5)+4 (n-1) \Big)\\
&  &
-\frac{\lambda  n (2 n+1) \left(\left(A^2-1\right) \lambda ^2+1\right)}{12 \left(n+\frac{1}{2}\right)^2} \Bigg\{ \frac{6 \lambda ^2 (2 n-1) (3-2 n)^2}{n \left(\lambda  \left(A^2 (-\lambda )+\lambda +2\right)+1\right)^2}\\
&  &
+\frac{12 \lambda  (1-2 n)^2}{n \left(\left(A^2-1\right) \lambda ^2-2 \lambda -1\right)}
+\frac{\lambda ^3 (2 n-1) (3-2 n)^2 (5-2 n)^2}{(n-1) n \left(\left(A^2-1\right) \lambda ^2-2 \lambda -1\right)^3}+24\Bigg\}\Bigg]
\end{eqnarray*} 
implies
\begin{eqnarray*}
&  &
(2 n+1)\lambda ^{2 n}\left(\lambda  \left(A^2 (-\lambda )+\lambda +2\right)+1\right)^{-n-\frac{5}{2}} {\mathcal F}(A,n;\lambda )\\
& = &
\frac{\Gamma(2n)}{[\Gamma(n+\frac{1}{2})]^2}\left( 4^{-2 n}-\frac{3  \left(4^{-2 n-1} (2 n-1) \left(2 A^2 (n-1)-1\right)\right)}{n-1}\lambda ^2+o\left(\lambda ^2\right)\right)\,.
\end{eqnarray*}
This completes the proof of the second  statement of the  lemma. The proof of the first statement of the lemma is straightforward. \qed

\subsection{The case of $\ell <1$ and $M=-\frac{1}{2}-n $, $n= 1 , 2,\ldots$}
\label{ss5.4}

Taking into account the scaling, we  from now on can  suppose $\varepsilon =1$.  
Assume $\ell <1$ and denote
\[
 K_1 \left(r,t; m ;1 \right)  
  := 
C(m,\ell)   \left( \left(1 +\tau \right)^2-A^2\right)^{\frac{1}{2}+n}   F \left(   -\frac{1}{2}-n ,   -\frac{1}{2}-n ;1;\frac{\left( \tau  -1\right)^2-A^2}{\left( \tau  +1\right)^2-A^2}\right), \quad C(m,\ell)\not=0  \,.  
\]
Consider the function
\[
\frac{\partial}{\partial t} K_1 \left(r,t; m ;1 \right)  
  := 
C(m,\ell) (1-\ell)t^{-\ell} \frac{\partial}{\partial \tau }  \left( \left(1 +\tau \right)^2-A^2\right)^{\frac{1}{2}+n}   F \left(   -\frac{1}{2}-n ,   -\frac{1}{2}-n ;1;\frac{\left( \tau  -1\right)^2-A^2}{\left( \tau  +1\right)^2-A^2}\right).  
\]
\begin{lemma}
In the asymptotic expansion of the function 
\begin{eqnarray*}
\frac{\partial}{\partial \tau }  \left( \left(1 +\tau \right)^2-A^2\right)^{\frac{1}{2}+n}   F \left(   -\frac{1}{2}-n ,   -\frac{1}{2}-n ;1;\frac{\left( \tau  -1\right)^2-A^2}{\left( \tau  +1\right)^2-A^2}\right)  
\end{eqnarray*}
as $\tau  \to \infty$ there is a term containing $A^2$:
\[
\tau  ^{2 n-2}\frac{ 2^{2 n+1} \left(2 A^2 n-1\right) \Gamma (n)}{\sqrt{\pi } \Gamma \left(n-\frac{1}{2}\right)}\,.   
\] 
\end{lemma}
\medskip

\noindent
{\bf Proof.} We consider the series about $\lambda =0$ after the change of variable $\lambda =\tau ^{-1}$:  
\[
\frac{\partial }{\partial \lambda }\left(\left(\left(\frac{1}{\lambda }+1\right)^2-A^2\right)^{n+\frac{1}{2}} F\left(-n-\frac{1}{2},-n-\frac{1}{2};1;\frac{\left(\frac{1}{\lambda }-1\right)^2-A^2}{\left(1+\frac{1}{\lambda }\right)^2-A^2}\right)\right)\,.
\]
We remind that   $\varphi \in C_0^\infty(0,1) $ implies $A^2=(1-\ell)^2r^2 >constant >0$.  Then,
\begin{eqnarray*}
& &\frac{\partial }{\partial \lambda }\left(\left(\left(\frac{1}{\lambda }+1\right)^2-A^2\right)^{n+\frac{1}{2}} F\left(-n-\frac{1}{2},-n-\frac{1}{2};1;\frac{\left(\frac{1}{\lambda }-1\right)^2-A^2}{\left(1+\frac{1}{\lambda }\right)^2-A^2}\right)\right)\\
& = &
-(2 n+1)\left((\lambda +1)^2-A^2 \lambda ^2\right)^{n-\frac{5}{2}}\lambda ^{-2 n-2}{\mathscr F}(\lambda ;n,A)\,,
\end{eqnarray*}
where
\begin{eqnarray}
\label{30}
&  &
{\mathscr F}(\lambda ;n,A)\\
& := &
(\lambda +1) \left(\lambda  \left(A^2 (-\lambda )+\lambda +2\right)+1\right)^2 F\left(-n-\frac{1}{2},-n-\frac{1}{2};1;1-\frac{4 \lambda }{(\lambda +1)^2-(A \lambda )^2}\right)\nonumber\\
&  &
+ \lambda  (2 n+1) \left(\left(A^2-1\right) \lambda ^2+1\right) \left((\lambda +1)^2-A^2 \lambda ^2\right)F\left(\frac{1}{2}-n,\frac{1}{2}-n;2;1-\frac{4 \lambda }{(\lambda +1)^2-(A \lambda )^2}\right)\,.\nonumber
\end{eqnarray}
We  consider  the function 
${\mathscr F}(\lambda ;n,A)$ around the point $\lambda =0 $.
In order to   substitute the following expansion    
\begin{eqnarray}
\label{29z}
&  &
z:=1-\frac{4 \lambda }{(\lambda +1)^2-(A \lambda )^2} =1-4 \lambda +8 \lambda ^2-4 \left(A^2+3\right) \lambda ^3+O\left(\lambda ^4\right)
\end{eqnarray}
in the series for the function ${\mathscr F}(\lambda ;n,A) $ we appeal to
 \cite[(12) Sec. 2.10]{B-E}:
\begin{eqnarray*}
F(a, b ; a+b+m ; z) \frac{1}{\Gamma(a+b+m)}
& = &
\frac{\Gamma(m)}{\Gamma(a+m) \Gamma(b+m)} \sum_{n=0}^{m-1} \frac{(a)_{n}(b)_{n}}{(1-m)_{n} n !}(1-z)^{n} \\
&  &
+\frac{(1-z)^{m}(-1)^{m}}{\Gamma(a) \Gamma(b)} \sum_{n=0}^{\infty} \frac{(a+m)_{n}(b+m)_{n}}{(n+m) ! n !}\left[h_{n}^{\prime \prime}-\ln (1-z)\right](1-z)^{n} \,,  \nonumber  
\end{eqnarray*}
where  $
-\pi<\arg (1-z)<\pi $, \quad $a, b, \neq 0,-1,2, \ldots $,    
\[
h_{n}^{\prime \prime}=\psi(n+1)+\psi(n+m+1)-\psi(a+n+m)-\psi(b+n+m), \nonumber
\]
and  the term $\sum_{n}^{m-1}$ in the expression for $ F(a, b ; a+b+m ; z)$  is to be interpreted as zero when $m=0$.
 
Hence  with $m=2n+2,\quad n=1,2,\ldots$, we obtain 
\begin{eqnarray*}
F\left(-n-\frac{1}{2},-n-\frac{1}{2};1;z\right) 
& = &
\frac{\Gamma(2n+2)}{[\Gamma(n+\frac{3}{2})]^2} \sum_{k=0}^{2n+1} \frac{[(-n-\frac{1}{2})_{k}]^2}{(-2n-1)_{k} k !}(1-z)^{k}+(1-z)^{2n+2}|\ln (1-z)|O(1)  \,,  \nonumber  
\end{eqnarray*}
and with $m=2n+1,\quad n=1,2,\ldots$, we derive 
\begin{eqnarray*}
F\left(\frac{1}{2}-n,\frac{1}{2}-n;2;z\right)
& = &
\frac{\Gamma(2n+1)}{[\Gamma(n+\frac{3}{2})]^2} \sum_{k=0}^{2n} \frac{[(\frac{1}{2}-n)_{k}]^2}{(-2n)_{k} k !}(1-z)^{k} 
+(1-z)^{2n+1}|\ln (1-z)O(1) \,.    
\end{eqnarray*}
We keep the terms of order $(1-z)^2 $ and  of order $(1-z)  $  in  the finite sums  
\begin{eqnarray*}
F\left(-n-\frac{1}{2},-n-\frac{1}{2};1;z\right) 
& = &
(2n+1) \frac{\Gamma(2n+1)}{[\Gamma(n+\frac{3}{2})]^2} \sum_{k=0}^{2} \frac{[(-n-\frac{1}{2})_{k}]^2}{(-2n-1)_{k} k !}(1-z)^{k}+(1-z)^{3}|\ln (1-z)|O(1)  \,,  \nonumber  
\end{eqnarray*}
and  
\begin{eqnarray*}
F\left(\frac{1}{2}-n,\frac{1}{2}-n;2;z\right)
& = &
\frac{\Gamma(2n+1)}{[\Gamma(n+\frac{3}{2})]^2} \sum_{k=0}^{1} \frac{[(\frac{1}{2}-n)_{k}]^2}{(-2n)_{k} k !}(1-z)^{k} 
+(1-z)^{2 }|\ln (1-z)|O(1) \,,     
\end{eqnarray*}
respectively. On the other hand    
\begin{eqnarray}
\label{30n}
&  &
(\lambda +1) \left(\lambda  \left(A^2 (-\lambda )+\lambda +2\right)+1\right)^2=1+5 \lambda +\left(10-2 A^2\right) \lambda ^2+\left(10-6 A^2\right) \lambda ^3+O\left(\lambda ^4\right)\,,\\
\label{31}
&  &
 \lambda  (2 n+1) \left(\left(A^2-1\right) \lambda ^2+1\right) \left((\lambda +1)^2-A^2 \lambda ^2\right)=  (2 n+1)\lambda+ (4 n+2)\lambda ^2+O\left(\lambda ^4\right) \,,
\end{eqnarray}
and
\begin{eqnarray}
\label{32}
\left((\lambda +1)^2-A^2 \lambda ^2\right)^{n-\frac{5}{2}}
& = &
1+  (2 n-5)\lambda-\frac{1}{2}  (2 n-5) \left(A^2-2 n+6\right)\lambda ^2+O\left(\lambda ^3\right)\,.
\end{eqnarray}
We substitute in $-(2 n+1)\left((\lambda +1)^2-A^2 \lambda ^2\right)^{n-\frac{5}{2}}{\mathscr F}(\lambda ;n,A)$ the  relations   (\ref{29z}), (\ref{30n}), (\ref{31}), (\ref{32})  and obtain
\[
-(2 n+1)\left((\lambda +1)^2-A^2 \lambda ^2\right)^{n-\frac{5}{2}}{\mathscr F}(\lambda ;n,A)= -\frac{(2 n+1) \Gamma (2 n+2)}{\Gamma \left(n+\frac{3}{2}\right)^2}+\frac{ 2^{2 n+1} \left(2 A^2 n-1\right) \Gamma (n)}{\sqrt{\pi } \Gamma \left(n-\frac{1}{2}\right)}\lambda ^2+O\left(\lambda ^3\right)\,.   
\]
The lemma is proved. \qed

\subsection{The case of $\ell <1$ and $M=\frac{1}{2} $}

Similar to what was done in the   previous subsection, we replace $\tau$ with $\lambda =\tau ^{-1}\to 0$ as $\tau \to \infty $.
\begin{lemma}
In the expansion for small $\lambda \searrow  0$ of the function 
\begin{eqnarray*}
& &
\frac{\partial }{\partial \lambda }\left(\left(\left(\frac{1}{\lambda }+1\right)^2-A^2\right)^{-\frac{1}{2}} F\left(\frac{1}{2},\frac{1}{2};1;\frac{\left(\frac{1}{\lambda }-1\right)^2-A^2}{\left(1+\frac{1}{\lambda }\right)^2-A^2}\right)\right)
\end{eqnarray*} 
there is a term $-(4 \pi)^{-1}\left(6 A^2+3\right) \lambda ^2\ln  (\lambda )
$, which is  depending on $A$.
\end{lemma}
\smallskip

\noindent
{\bf Proof.} 
Indeed,
\[
\frac{\partial }{\partial \lambda }\left(\left(\left(\frac{1}{\lambda }+1\right)^2-A^2\right)^{-\frac{1}{2}} F\left(\frac{1}{2},\frac{1}{2};1;\frac{\left(\frac{1}{\lambda }-1\right)^2-A^2}{\left(1+\frac{1}{\lambda }\right)^2-A^2}\right)\right)\\
  =  
\left( \left(1-A^2 \right) \lambda ^2+2 \lambda +1\right)^{-\frac{5}{2}}{\mathscr F} (A,\lambda )\,,
\]
where
\begin{eqnarray}
\label{30new}
{\mathscr F} (A,\lambda )
& := &\Bigg\{(-A^2 \lambda ^3-A^2 \lambda ^2+\lambda ^3+3 \lambda ^2+3 \lambda +1 )F\left(\frac{1}{2},\frac{1}{2};1;\frac{\left(\frac{1}{\lambda }-1\right)^2-A^2}{\left(1+\frac{1}{\lambda }\right)^2-A^2}\right) \nonumber\\
&  &
- (A^2 \lambda ^3-\lambda ^3+\lambda)F\left(\frac{3}{2},\frac{3}{2};2;\frac{\left(\frac{1}{\lambda }-1\right)^2-A^2}{\left(1+\frac{1}{\lambda }\right)^2-A^2}\right) \Bigg\}\,.
\end{eqnarray}
Then it is easy to see the following asymptotic expansions 
\begin{eqnarray}
\left(\left(1-A^2\right) \lambda ^2+2 \lambda +1\right)^{-\frac{5}{2}}
& = &
1-5 \lambda +\frac{5}{2} \left(A^2+6\right) \lambda ^2-\frac{35}{2} \left(A^2+2\right) \lambda ^3+O\left(\lambda ^4\right)\,, \nonumber \\
\label{41}
1-\frac{\left(\frac{1}{\lambda }-1\right)^2-A^2}{\left(\frac{1}{\lambda }+1\right)^2-A^2}
& = &
4 \lambda -8 \lambda ^2+4 \left(A^2+3\right) \lambda ^3+O\left(\lambda ^4\right)\,.
\end{eqnarray} 
Hence, according to (\ref{F12exp}) and (\ref{F32exp}), we have
\[
F\left(\frac{1}{2},\frac{1}{2};1;\frac{\left(\frac{1}{\lambda }-1\right)^2-A^2}{\left(\frac{1}{\lambda }+1\right)^2-A^2}\right)\\
=
\frac{2 \ln(2)-\ln  \left(\lambda \right)}{\pi }+\frac{2 \ln(2)-\ln  \left(\lambda \right)}{\pi }\lambda+\frac{-4 A^2-9+2 \ln(2)-\ln  \left(\lambda \right)}{4 \pi }\lambda ^2+O\left(\lambda ^3\right)
\]
and 
\begin{eqnarray*}
F\left(\frac{3}{2},\frac{3}{2};2;\frac{\left(\frac{1}{\lambda }-1\right)^2-A^2}{\left(\frac{1}{\lambda }+1\right)^2-A^2}\right)
& = &
\frac{1}{\pi  \lambda }+\frac{5-2 \ln  (2)+\ln  (\lambda )}{\pi }
+\frac{-4 A^2+47-18 \ln  (4)+18 \ln  (\lambda )}{4 \pi }\lambda \\
&  &
+\frac{4 A^2+84-39 \ln  (4)+39 \ln  (\lambda )}{4 \pi }\lambda ^2+O\left(\lambda ^3\right)\,.
\end{eqnarray*} 
It remains to substitute obtained expansions into (\ref{30new}) to derive 
\begin{eqnarray*}
{\mathscr F} (A,\lambda )
& = &
\frac{-1+\ln  (4)-\ln  (\lambda )}{\pi }+\frac{5  (-1+\ln  (4)-\ln  (\lambda ))}{\pi }\lambda \\
&  &
+\frac{ -4 A^2-4 A^2 \ln  (4)-52+86 \ln  (2)-43 \ln  (\lambda )+4 A^2 \ln  (\lambda )}{4 \pi }\lambda ^2+O\left(\lambda ^3\right)\,.
\end{eqnarray*}
Thus,
\begin{eqnarray*}
& &
\frac{\partial }{\partial \lambda }\left(\left(\left(\frac{1}{\lambda }+1\right)^2-A^2\right)^{-\frac{1}{2}} F\left(\frac{1}{2},\frac{1}{2};1;\frac{\left(\frac{1}{\lambda }-1\right)^2-A^2}{\left(1+\frac{1}{\lambda }\right)^2-A^2}\right)\right)\\
& = &
\Bigg\{1-5 \lambda +\frac{5}{2} \left(A^2+6\right) \lambda ^2-\frac{35}{2} \left(A^2+2\right) \lambda ^3+O\left(\lambda ^4\right)\Bigg\}\\
&  &
\times \Bigg\{\frac{-1+\ln  (4)-\ln  (\lambda )}{\pi }+\frac{5  (-1+\ln  (4)-\ln  (\lambda ))}{\pi }\lambda \\
&  &
+\frac{ -4 A^2-4 A^2 \ln  (4)-52+86 \ln  (2)-43 \ln  (\lambda )+4 A^2 \ln  (\lambda )}{4 \pi }\lambda ^2+O\left(\lambda ^3\right)\Bigg\}\,.
\end{eqnarray*} 
Finally, we obtain
\begin{eqnarray*}
& &
\frac{\partial }{\partial \lambda }\left(\left(\left(\frac{1}{\lambda }+1\right)^2-A^2\right)^{-\frac{1}{2}} F\left(\frac{1}{2},\frac{1}{2};1;\frac{\left(\frac{1}{\lambda }-1\right)^2-A^2}{\left(1+\frac{1}{\lambda }\right)^2-A^2}\right)\right)\\
& = &
\frac{-\ln  (\lambda )-1+\ln  (4)}{\pi }+\frac{-\left(6 A^2+3\right) \ln  (\lambda )-14 A^2+12 A^2 \ln  (2)-12+6\ln  (2)}{4 \pi }\lambda ^2+O\left(\lambda ^3\right)\,.
\end{eqnarray*} 
The lemma is proved. \qed

\subsection{The case of $\ell <1$ and $M=-\frac{1}{2} $}
We appeal to the arguments  which  have been used in subsection~\ref{ss5.4} and consider  around the point $\lambda =0 $ the function  
$\left((\lambda +1)^2-A^2 \lambda ^2\right)^{-\frac{5}{2}}{\mathscr F}(\lambda ;0,A)$,
where $ {\mathscr F}(\lambda ;0,A)$ is defined in  (\ref{30}).
\begin{lemma}
In the expansion of the function  (\ref{30}) for small $\lambda \searrow  0$  there is a term   $(-2 A^2 -3+4\ln  (2))/\pi  $ 
depending on $A$.
\end{lemma}
\medskip

\noindent
{\bf Proof.} Indeed,  we have
\[
\frac{\partial }{\partial \lambda }\left(\left(\left(\frac{1}{\lambda }+1\right)^2-A^2 \right)^{1/2} F\left(-\frac{1}{2},-\frac{1}{2};1;\frac{\left(\frac{1}{\lambda }-1\right)^2-A^2}{\left(1+\frac{1}{\lambda }\right)^2-A^2}\right)\right) 
  =  
-\frac{1}{\lambda ^2 \left(-A^2 \lambda ^2+\lambda ^2+2 \lambda +1\right)^{3/2}}{\mathscr F}(\lambda ,A)\,,
\] 
where
\begin{eqnarray*}
{\mathscr F}(\lambda ,A)
&:= &\Bigg\{\left(-A^2 \lambda ^3-A^2 \lambda ^2+\lambda ^3+3 \lambda ^2+3 \lambda +1\right)F\left(-\frac{1}{2},-\frac{1}{2};1;\frac{\left(\frac{1}{\lambda }-1\right)^2-A^2}{\left(1+\frac{1}{\lambda }\right)^2-A^2}\right)\\
&  &
+\left(A^2 \lambda ^3-\lambda ^3+\lambda\right) F\left(\frac{1}{2},\frac{1}{2};2;\frac{\left(\frac{1}{\lambda }-1\right)^2-A^2}{\left(1+\frac{1}{\lambda }\right)^2-A^2}\right) \Bigg\}\,.
\end{eqnarray*}
Furthermore, according to subsection~\ref{ss5.4}, we can write 
\begin{eqnarray*}
F\left(-\frac{1}{2},-\frac{1}{2};1;z\right)
&= &
\frac{4}{\pi }+\frac{z-1}{\pi }+\frac{-2 \ln  (1-z)-5+8\ln  (2)}{16 \pi }(z-1)^2 \\
&  &
-\frac{3 (-2 \ln  (1-z)-5+8 \ln  (2))}{64 \pi }(z-1)^3 +O\left((z-1)^4\right)
\end{eqnarray*}
and
\begin{eqnarray*}
F\left( \frac{1}{2},  \frac{1}{2};2;z\right)
&= &
\frac{4}{\pi }+\frac{4\ln (2)-\ln  (1-z )-3}{\pi }(z-1)-\frac{3 (-6 \ln  (1-z)-17+24 \ln  (2))}{16 \pi }(z-1)^2 \\
&  &
+\frac{15 (-5 \ln  (1-z)-14+20 \ln  (2))}{64 \pi } (z-1)^3+O\left((z-1)^4\right)\,.
\end{eqnarray*}
We substitute  the   relation (\ref{41}) in the expansions for $F\left( -\frac{1}{2},  -\frac{1}{2};1;z\right)$ and $F\left( \frac{1}{2},  \frac{1}{2};2;z\right)$ and derive 
\begin{eqnarray*}
&  &
F\left(-\frac{1}{2},-\frac{1}{2};1;\frac{\left(\frac{1}{\lambda }-1\right)^2-A^2}{\left(\frac{1}{\lambda }+1\right)^2-A^2}\right)\\
& = & 
\frac{4}{\pi }-\frac{4}{\pi } \lambda +\frac{-2 \ln  (\lambda )+3+4\ln  (2)}{\pi }\lambda ^2+\frac{-4 A^2+2 \ln  (\lambda )-3-4\ln  (2)}{\pi }\lambda ^3+O\left(\lambda ^4\right)\,,\\
&  &
F\left(\frac{1}{2},\frac{1}{2};2;\frac{\left(\frac{1}{\lambda }-1\right)^2-A^2}{\left(\frac{1}{\lambda }+1\right)^2-A^2}\right)\\
& = &
\frac{4}{\pi }-\frac{4 (-\ln  (\lambda )-3+2\ln  (2))}{\pi }\lambda  +\frac{10 \ln  (\lambda )+19-20 \ln  (2)}{\pi }\lambda ^2\\
&  &
+\frac{ 4 A^2 \ln  (\lambda )+16 A^2-8 A^2 \ln  (2)+15 \ln  (\lambda )+26-30 \ln  (2)}{\pi }\lambda ^3+O\left(\lambda ^4\right)\,.
\end{eqnarray*}
On the other hand,
\begin{eqnarray*}
-\frac{1}{\lambda ^2 \left(-A^2 \lambda ^2+\lambda ^2+2 \lambda +1\right)^{3/2}}
& = &
-\frac{1}{\lambda ^2}+\frac{3}{\lambda }-\frac{3}{2} \left(A^2+4\right)+\frac{5}{2} \left(3 A^2+4\right) \lambda -\frac{15}{8} \left(A^4+12 A^2+8\right) \lambda ^2\\
&  &
+\frac{21}{8} \left(5 \left(A^2+4\right) A^2+8\right) \lambda ^3+O\left(\lambda ^4\right)\,.
\end{eqnarray*} 
Thus, we obtain
\begin{eqnarray*}
& &
\frac{\partial }{\partial \lambda }\left(\left(\left(\frac{1}{\lambda }+1\right)^2-A^2\right)^{1/2} F\left(-\frac{1}{2},-\frac{1}{2};1;\frac{\left(\frac{1}{\lambda }-1\right)^2-A^2}{\left(1+\frac{1}{\lambda }\right)^2-A^2}\right)\right)\\
& = &
-\frac{4}{\pi  \lambda ^2}+\frac{-2 A^2-3+4\ln  (2)-2 \ln  (\lambda )}{\pi }+O\left(\lambda ^2\right)\,.
\end{eqnarray*} 
The lemma is proved. \qed
\medskip

The Huygens' principle's is a local property, that is, it can be verified for the small time. On the other hand, the proof of the necessity part presented in this paper is based on the large time asymptotics. In fact, by taking into account the scaling invariance of the operator (\ref{DO}), the proof   can be easily modified  to the small time by the implementing a large auxiliary parameter. In this paper, we avoid that  modification,   which   creates  an  unnecessary cumbersome, even though
 it could be  a useful  tool for the further generalizations.

\end{document}